\documentclass[aps,prd,amsmath,amssymb,preprintnumbers,nofootinbib,a4paper,11pt]{revtex4-1}
\pdfoutput=1
\usepackage{amsthm}
\usepackage{graphicx}
\usepackage{url}
\usepackage{color}
	\definecolor{rossoCP3}{cmyk}{0,.88,.77,.40}
		\definecolor{graa}{rgb}{0.8,0.8,0.8}
		\definecolor{blaa}{rgb}{0.2,0.2,0.6}
\usepackage{dcolumn}% Align table columns on decimal point
\usepackage{bm}% bold math
\usepackage{bbm}
\usepackage{pxfonts}

\usepackage{epsfig}
\usepackage{placeins}

\usepackage[ margin=5pt, font=small,labelfont=bf, justification=raggedright]{caption}
\usepackage{youngtab}
\usepackage{slashed}
\usepackage{subfig}

\newcommand{\beq}{\begin{eqnarray}}
\newcommand{\eeq}{\end{eqnarray}}

\newcommand{\bmp}{\noindent\begin{minipage}{16cm}}
\newcommand{\emp}{\end{minipage}\vskip 7mm} % 7mm untightened

\def\lsim{\mathrel{\rlap{\lower4pt\hbox{\hskip1pt$\sim$}}
    \raise1pt\hbox{$<$}}}                % less than or approx. symbol
\def\gsim{\mathrel{\rlap{\lower4pt\hbox{\hskip1pt$\sim$}}
    \raise1pt\hbox{$>$}}}                % greater than or approx. symbol

\begin{document}
%%%%%%%%%%%%%%%%%%%%%%%%%%%%%%%%%%%%%%%%%%%%%%%%%%%%%%%%%%%%%%%%%%%%%%%%%%%
\title{\Large  \color{rossoCP3}  125 GeV Higgs from a not so light Technicolor Scalar }
\author{Roshan Foadi$^{\color{rossoCP3}{\clubsuit}}$}%\email{foadi@}
\author{Mads T. Frandsen$^{\color{rossoCP3}{\varheartsuit}}$}%\email{frandsen@cp3-origins.net}
\author{Francesco Sannino$^{\color{rossoCP3}{\varheartsuit}}$}%\email{sannino@cp3-origins.net}

\affiliation{$^{\color{rossoCP3}{\clubsuit}}$ {Centre for Cosmology, Particle Physics and Phenomenology (CP3)
Chemin du Cyclotron 2 \\  Universit\'e catholique de Louvain, Belgium}}
 \affiliation{
$^{\color{rossoCP3}{\varheartsuit}}${ \color{rossoCP3}  \rm CP}$^{\color{rossoCP3} \bf 3}${\color{rossoCP3}\rm-Origins} \& the Danish Institute for Advanced Study {\color{rossoCP3} \rm DIAS},\\
University of Southern Denmark, Campusvej 55, DK-5230 Odense M, Denmark.
}
\begin{abstract}
Assuming that the observed Higgs-like resonance at the Large Hadron Collider is a technicolor isosinglet scalar (the technicolor Higgs), we argue that the standard model top-induced radiative corrections reduce its mass towards the desired experimental value. We discuss conditions for the spectrum of technicolor theories to feature a technicolor Higgs with the phenomenologically required dynamical mass. We consider different representations under the technicolor gauge group, and employ scaling laws in terms of the dimension of the representation. We also summarize the potential effects of walking dynamics on the mass of the technicolor Higgs.
 \\
[.1cm]
{\footnotesize  \it Preprint: CP$^3$-Origins-2012-028 \& DIAS-2012-29}
 \end{abstract}

\maketitle

\section{Introduction}
Recently the ATLAS and CMS collaborations have announced the discovery of a new boson with the approximate mass of 125 GeV~\cite{:2012gk,:2012gu}.
The observed decays to standard model (SM) di-boson pairs, $\gamma \gamma$, $Z Z^*$ and $WW^*$, are in rough agreement with
those expected from the SM Higgs. Here we assume this state to be a scalar, as suggested by the observed decays into $Z Z^*$ and $WW^*$~\cite{Frandsen:2012rj,Coleppa:2012eh,Eichten:2012qb}.

If new strong dynamics similar to technicolor (TC)~\cite{Weinberg:1975gm,Dimopoulos:1979es,Eichten:1979ah,Sannino:2008ha,Sannino:2009za} underlies the Higgs mechanism, is it possible for the corresponding spectrum to feature a 125 GeV composite scalar? To answer this question we must disentangle the SM radiative corrections from the dynamical mass $M_H^{TC}$, {\em i.e.} the mass stemming purely from composite dynamics. Because of the large and negative radiative corrections from the top loop, we will argue that for a SM-like top-Yukawa coupling the dynamical mass of the scalar required to match the observations is of the order of $M_H^{TC}\sim 600/\sqrt{N_{\rm TD}}$ GeV, where $N_{\rm TD}$ is the number of weak technidoublets. This is a significant increase in the value of the dynamical mass compared to the observed 125 GeV, often naively identified with the dynamical mass.  Additionally, if the dominant decay channels are into SM states, then for a physical mass of 125 GeV the TC Higgs will be narrow simply because of kinematics. A similar example in strong dynamics is the $f_0(980)$ resonance, which is extremely narrow because the decay mode into $K\bar{K}$ is below threshold \cite{Harada:1995dc}.

Thus the question we should ask ourselves is: Are there TC theories featuring a scalar singlet with dynamical mass $M_H^{TC}\sim 600/\sqrt{N_{\rm TD}}$ GeV? Scaling up two-flavor QCD gives an estimate for the lightest scalar singlet in the $1.0\ {\rm TeV}\lesssim M_H^{TC}\lesssim 1.4\ {\rm TeV}$ range, somewhat heavier than the required value. However we shall see that higher-dimensional technifermion representations, in $SU(N_{\rm TC})$ TC, lead to a lighter TC-Higgs mass as $N_{\rm TC}$ grows, and that the required value for $M_H^{TC}$ is already attained for relatively small values of $N_{\rm TC}$.

Furthermore, reduction of the TC-Higgs dynamical mass may originate from walking (or near-conformal) dynamics~\cite{Holdom:1981rm}, both for the fundamental or for higher-dimensional representations. Walking dynamics is useful to alleviate the tension with flavor changing neutral currents, and to reduce the value of the $S$-parameter~\cite{Appelquist:1998xf,Kurachi:2006ej}. The latter, however, is expected to vanish neither in conformal field theories~\cite{Sannino:2010ca,Sannino:2010fh,DiChiara:2010xb}, nor in TC theories with near-conformal dynamics~\cite{Foadi:2012ga}. In the literature a light TC Higgs originating from walking-type dynamics is also known as {\em technidilaton}~\cite{Yamawaki:1985zg,Bando:1986bg,Dietrich:2005jn,Appelquist:2010gy}.

The remainder of this paper is organized as follows. In Sec.~\ref{sec:loop} we set up an effective Lagrangian including the TC Higgs, the SM particles, and their interactions. We then  compute the radiative corrections from the top quark and the SM gauge bosons to the TC-Higgs mass. In Sec.~\ref{Smass} we analyze the scaling of the dynamical TC-Higgs mass as a function of the dimension of the representation, $d(R_{\rm TC})$, and the number of technidoublets $N_{\rm TD}$, and show -- taking into account the SM radiative corrections -- which TC theories can give a physical TC-Higgs mass of 125 GeV. We then turn to the possibility of alternative or additional reductions of the dynamical mass from walking dynamics. We argue that walking TC can accommodate a 125 GeV Higgs even for small values of $d(R_{\rm TC})$ and $N_{\rm TD}$. Finally in Sec.~\ref{sec:conclusions} we offer our conclusions.
\section{Radiative corrections to the mass of the TC Higgs}\label{sec:loop}
We consider TC theories featuring, at scales below the mass of the technirho $M_{\rho}$, only the eaten Goldstone bosons and the TC Higgs $H$. We assume the TC dynamics to respect $SU(2)_c$ custodial isospin symmetry, and adopt a nonlinear realization for the composite states. The latter are thus classified according to linear multiplets of $SU(2)_c$: the electroweak Goldstone bosons $\pi^a$, with $a=1,2,3$, form an $SU(2)_c$ triplet, whereas the TC Higgs is an $SU(2)_c$ singlet. The elementary SM fields are linear multiplets of the electroweak group. The Yukawa interactions of the TC Higgs with SM fermions are induced by interactions beyond the TC theory itself, {\em e.g.} extended technicolor (ETC)~\cite{Dimopoulos:1979es,Eichten:1979ah}.

\begin{figure}
\includegraphics[width=12cm]{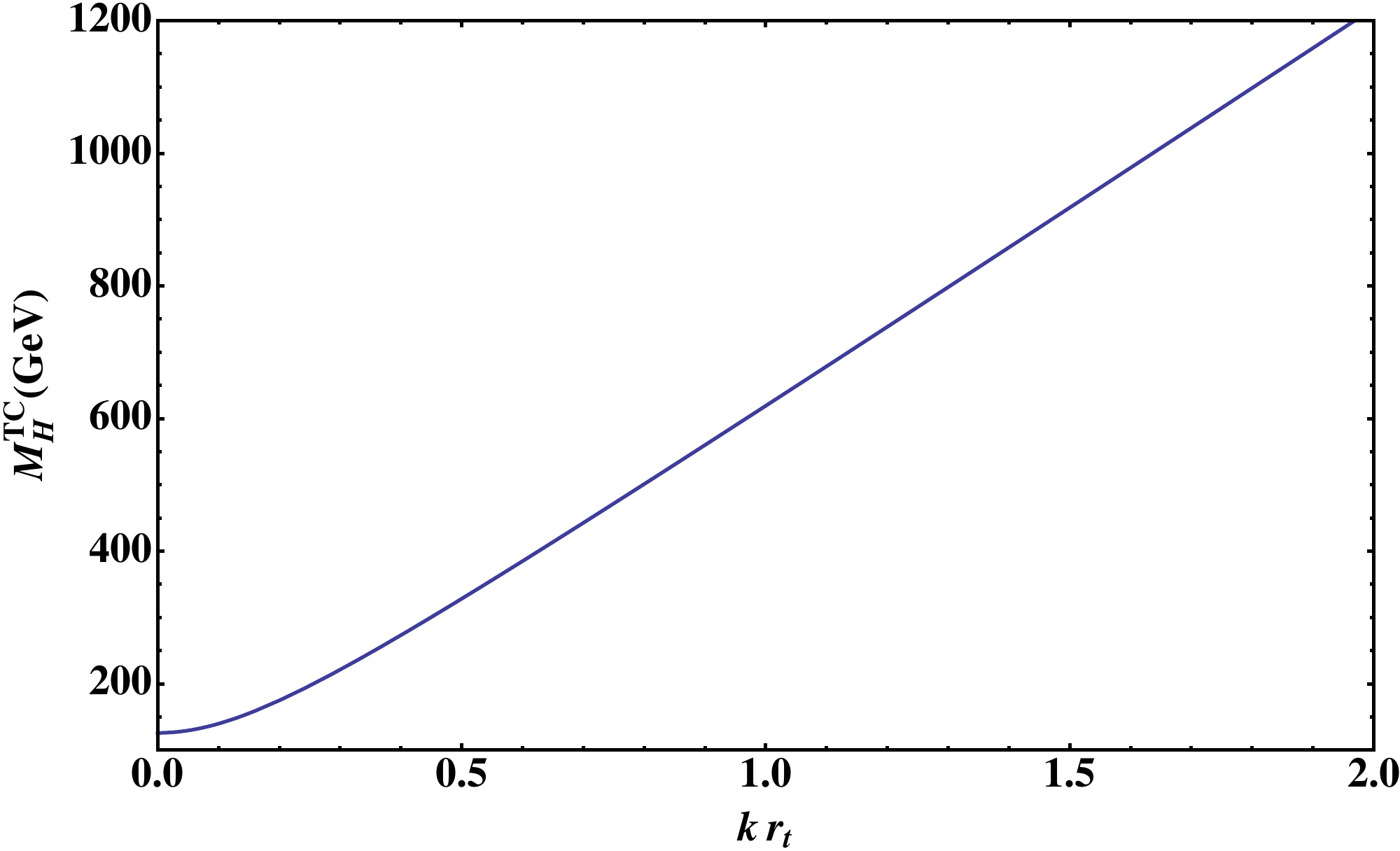}
\caption{Mass of the TC Higgs as a function of the product $\kappa \, r_t$, using the formula $M_H^{TC} = \sqrt{M^2_H + 12 \kappa^2 r_t^2 m_t^2}$. The latter is obtained from~\eqref{quadcorMrho} by neglecting the weak gauge boson contributions, and with the counterterm set to zero.}
\label{krt2}
\end{figure}

Assuming that the only non-negligible sources of custodial isospin violation are due to the Yukawa interactions, and retaining only the leading order terms in a momentum expansions, leads to the effective Lagrangian
\begin{eqnarray}
{\cal L} &=& {\cal L}_{\overline{\rm SM}}
+\left(1+\frac{2 r_\pi}{v}H+\frac{s_\pi}{v^2}H^2\right)\frac{v^2}{4}{\rm Tr}\ D_\mu U^\dagger D^\mu U
+ \frac{1}{2}\ \partial_\mu H\ \partial^\mu H - V[H]
 \nonumber \\
&-&
\ m_t\left(1+\frac{r_t}{v}H\right)
\Bigg[\overline{q}_L\ U\ \Bigg(\frac{1}{2}+T^3\Bigg)\ q_R + {\rm h.c.} \Bigg] \nonumber \\
&-& m_b\left(1+\frac{r_b}{v}H\right)
\Bigg[\overline{q}_L\ U\ \Bigg(\frac{1}{2}-T^3\Bigg) \ q_R + {\rm h.c.} \Bigg] + \cdots \nonumber \\
&-&\Delta S\ W^a_{\mu\nu}B^{\mu\nu}\ {\rm Tr}\  T^a U T^3 U^\dagger +{\cal O}\left(\frac{1}{M_\rho}\right)
\label{eq:L}
\end{eqnarray}
where $ {\cal L}_{\overline{\rm SM}}$ is the SM Lagrangian without Higgs and Yukawa sectors, the ellipses denote Yukawa interactions for SM fermions other than the top-bottom doublet $q\equiv(t,b)$, and ${\cal O}(1/M_\rho)$ includes higher-dimensional operators, which are suppressed by powers of $1/M_\rho$. In this Lagrangian $v\simeq 246$ GeV is the electroweak vev, $U$ is the usual exponential map of the Goldstone bosons produced by the breaking of the electroweak symmetry, $U=\exp\Big(i 2 \pi^a T^a/v\Big)$, with covariant derivative $D_\mu U\equiv \partial_\mu U -i g W^a_\mu T^a U + i g^\prime U B_\mu T^3$, $2T^a$ are the Pauli matrices, with $a=1,2,3$, and $V[H]$ is the TC Higgs potential. $\Delta S$ is the contribution to the $S$ parameter from the physics at the cutoff scale, and is assumed to vanish in the $M_\rho\to\infty$ limit. The interactions contributing to the Higgs self-energy are
\begin{eqnarray}
{\cal L}_H &\supset&  \frac{2\ m_W^2\ r_\pi}{v}\ H\ W^+_\mu\ W^{-\mu}
+\frac{m_Z^2\ r_\pi}{v}\ H\ Z_\mu\ Z^\mu - \frac{m_t\ r_t}{v}\ H\ \bar{t}\ t \nonumber \\
&+& \frac{m_W^2\ s_\pi}{v^2}\ H^2\ W^+_\mu\ W^{-\mu}
+ \frac{m_Z^2\ s_\pi}{2\ v^2}\ H^2\ Z_\mu\ Z^\mu\ .
\label{eq:hLagr}
\end{eqnarray}
The tree-level SM is recovered for
\begin{equation}
r_\pi =s_\pi=r_t=r_b=1  \ .
\end{equation}
\begin{figure}
\includegraphics[width=5.0in]{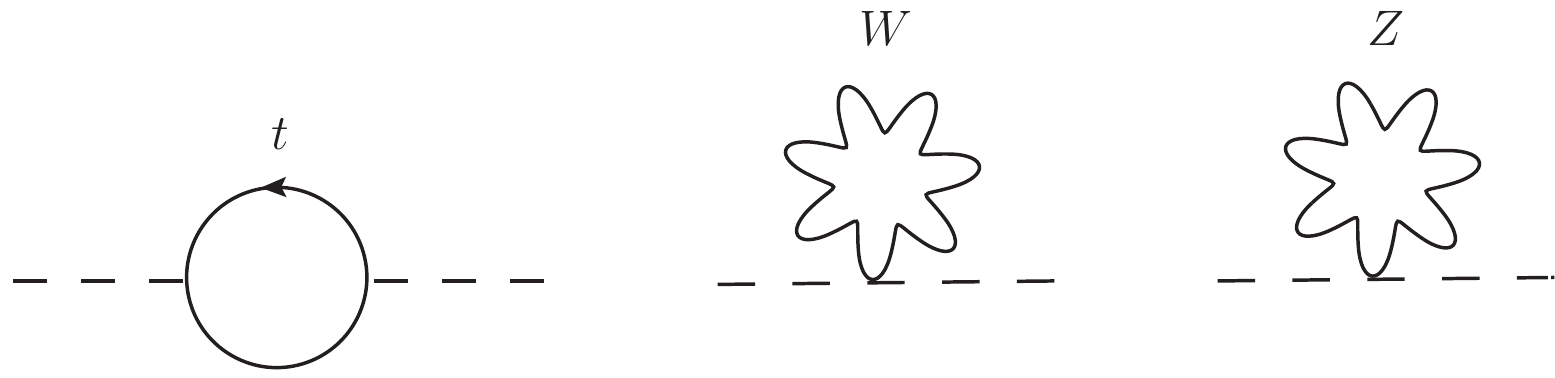}
\caption{Quadratically divergent diagrams contributing to the Higgs mass, with the interaction vertices given by (\ref{eq:hLagr}). The gauge boson exchanges are computed in Landau gauge: then the seagull diagrams, with a single $W$ and $Z$ exchange, are the only quadratically divergent one-loop diagrams with gauge boson exchanges.}
\label{fig:diagrams}
\end{figure}
We divide the radiative corrections to the TC Higgs mass into two classes: external contributions, corresponding to loop corrections involving elementary SM fields, and TC contributions, corresponding to loop corrections involving TC composites only. The latter contribute to the dynamical mass $M_H^{TC}$, whose size will be estimated in the next section by non-perturbative analysis. In order to isolate the SM contributions we work in Landau gauge. Here transversely polarized gauge boson propagators correspond to elementary fields, and massless Goldstone boson propagators correspond to TC composites. The only SM contributions to the TC Higgs mass which are quadratically divergent  in the cutoff   come from the diagrams of Fig.~\ref{fig:diagrams}. Retaining only the quadratically divergent terms leads to a physical mass $M_H$ given by
\begin{eqnarray}
M_H^2=(M_H^{TC})^2 + \frac{3 (4\pi \kappa F_{\Pi})^2 }{16\pi^2 v^2}
\left[-4 r_t^2 m_t^2 +2 s_\pi \left(m_W^2+\frac{m_Z^2}{2} \right) \right] + \Delta_{M_H^2}(4\pi \kappa F_{\Pi})\ ,
\label{quadcorMrho}
\end{eqnarray}
where $F_\Pi$ is the TC-pion decay constant, and $\Delta_{M_H^2}(4\pi \kappa F_{\Pi})$ is the counterterm. The cutoff is estimated to be $4\pi\kappa F_\pi$, where $\kappa$ is a number of order one. The latter scales like $1/\sqrt{d(R_{\rm TC})}$ if the cutoff is identified with the technirho mass, or is a constant if the cutoff is of the order of $4\pi F_{\Pi}$. Provided $r_t$ is also of order one, the dominant radiative correction is due to the top quark. For instance, if $F_\Pi=v$, which is appropriate for a TC theory with one weak technidoublet, then $\delta M_H^2 \sim - 12 \kappa^2 r_t^2 m_t^2 \sim -\kappa^2 r_t^2 (600 \, {\rm GeV})^2$. In Fig.~\ref{krt2} we plot the mass of the TC Higgs as a function of the product $\kappa \, r_t$ using the formula $M_H^{TC} = \sqrt{M^2_H + 12 \kappa^2 r_t^2 m_t^2}$. This is obtained in the simple approximation of neglecting the weak gauge boson contributions, and having set to zero the counterterm in~\eqref{quadcorMrho}. This shows that the dynamical mass of the TC Higgs can be substantially heavier than the physical mass, $M_H\simeq 125$ GeV.
\section{Estimates of the TC Higgs  couplings to Gauge Bosons and Fermions}\label{sec:couplings}
In the previous section we used an effective Lagrangian approach to disentangle the SM radiative corrections from the intrinsic value of the  TC-Higgs mass stemming from the underlying pure TC dynamics. In this section we discuss the origin and size of the relevant effective couplings of the TC-Higgs to the SM fields. The natural values of these couplings are those used for the estimate above. In TC the couplings to the SM vector bosons and fermions have different origins and we discuss them in turn.
\subsection{Couplings to SM gauge bosons}\label{sec:}
The couplings of a TC-Higgs to SM vector bosons have been studied in e.g. \cite{Belyaev:2008yj,Foadi:2008xj,Frandsen:2012rj}. 
To see that these couplings are expected to be similar to those of the SM Higgs, consider the relevant SM Lagrangian 
\begin{eqnarray}
\mathcal{L}=-\frac{1}{v}(h\partial_\mu w \cdot \partial_\mu w) \ ,
\end{eqnarray} 
where the $w$ fields correspond to the longitudinal component of the massive SM gauge bosons. A similar Lagrangian term is used to describe the decay of the QCD $f_0(500)$   (also known as  the $\sigma$) state into pions. It is sufficient to replace $\omega \to \pi$ and $1/v \to g_{\sigma \pi \pi}$. To determine the overall coefficient of this operator a fit of this coupling to pion-pion scattering data has been performed in e.g. \cite{Harada:1995dc} finding that $g_{\sigma \pi \pi} \sim 1/200 $ MeV$^{-1}$ which is of the order of $1/f_\pi$.  
Therefore in the simplest TC models where the technipion decay constant is identified with the electroweak vev $v$, we find that the coupling of the TC Higgs to the longitudinal components of the SM gauge bosons is indeed of the order $1/v$. In particular, when gauging the TC theory under the electroweak interactions we then find, as in the SM, $r_\pi \sim 1$.
\subsection{Couplings to fermions}\label{sec:}
The couplings of a TC Higgs to SM fermions arise from interactions beyond TC. Several extensions of TC have been suggested in the literature to address the problem of fermion mass generation. Some
of the extensions use additional strongly coupled gauge dynamics \cite{Dimopoulos:1979es,Eichten:1979ah} others introduce fundamental
scalars \cite{Samuel:1990dq}. Many variants of these schemes exist and a review of some are given in \cite{Hill:1994hp}. 

Here we assume that the couplings to SM fermions arise from an unspecified ETC model\footnote{Attempts towards a realistic ETC model can be found in e.g. \cite{Appelquist:2003hn, Appelquist:2004ai} where three different ETC scales are invoked to explain the fermion generation mass hierarchies and it is assumed that  $\gamma_m \sim 1$. The authors start from an $SU(5)_{\rm ETC}$ gauge group that commutes with the SM gauge group. The ETC gauge group breaks consecutively down to $SU(2)_{\rm TC}$ in three stages.} yielding, at the electroweak scale, the following relevant four fermion operator 
\begin{eqnarray}
\mathcal{L}_{ETC} =  g_{\rm ETC}^2 \frac{\bar{Q}Q \bar{f}f}{\Lambda_{\rm ETC}^2} 
+ ...
\end{eqnarray}
We restrict to the top quark and have:
\begin{eqnarray}
g_{\rm ETC}^2 \frac{\bar{Q}Q \bar{f}f}{\Lambda_{\rm ETC}^2} \to 
g_{\rm ETC}^2 \frac{\langle \bar{Q}Q \rangle_{\rm TC}  (r_t H+v)\bar{f}f }{\Lambda_{\rm ETC}^2 v}  = 
\frac{ m_f }{v} (v \bar{f}f +  r_t H \bar{f}f )
\ .
\end{eqnarray}
Here we are evaluating the technifermion condensate at the electroweak scale and $r_t$ naturally represents the overlap of the scalar state with the fermion-antitermion TC operator. The requirement that this 4-fermion operator provides the full SM fermion mass implies that the TC Higgs Yukawa interactions will be proportional to $m_f$ as it is for the SM Higgs interactions.  In a strongly coupled theory with a single scale $r_t$ must be of order unity.

We can compute $r_t$ explicitly when the TC Higgs state $H$ is identified with a Techni-dilaton (TD)  because one has $\bar{Q}Q = \langle \bar{Q}Q \rangle e^{\frac{-(3-\gamma_m) H}{F_{TD}}} U$ , 
with $F_{TD}$ the TD decay constant and $\gamma_m$ the anomalous dimension of the technifermion mass operator, e.g. \cite{Bando:1986bg,Hashimoto:2011ma,Matsuzaki:2012vc}.  
Therefore $r_t=(3-\gamma_m) \frac{v}{F_{TD}}$ which confirms our expectation that  when the TD scale $F_{TD}$ approaches the electroweak scale $F_\Pi=v$ we have $r_t$ of order unity, assuming the physical values $ 0 \leq \gamma_m < 2$. The SM Higgs coupling is recovered for $\gamma_m=2$ as it should.

\section{The dynamical mass of the TC Higgs}\label{Smass}
In QCD the lightest scalar is the $\sigma$ meson (also termed $f_0(500)$ in PDG), with a measured mass between 400 and 550 MeV~\cite{Beringer:1900zz} in agreement with early determinations \cite{Harada:1995dc}. {Scaling up two-flavor QCD yields a TC Higgs dynamical mass in the $1.0\ {\rm TeV}\lesssim M_H^{TC}\lesssim 1.4\ {\rm TeV}$ range.  This estimate changes when considering TC theories which are not an exact replica of two-flavor QCD. Here we consider $SU(N_{\rm TC})$ gauge theories, and determine the {\it geometric scaling} of the TC-Higgs dynamical mass, {\em i.e.} the value of $M_H^{TC}$, as function of $N_{\rm TC}$, the dimension of the TC-matter representation, $d(R_{\rm TC})$, and the number of weak technidoublets $N_{\rm TD}$. For a generalization to gauge groups other than $SU(N_{\rm TC})$ see~\cite{Sannino:2009aw,Mojaza:2012zd}.} We then discuss possible effects of walking dynamics on $M_H^{TC}$, which are not automatically included in the geometric scaling. Taking into account the SM induced radiative corrections discussed in Sec.~\ref{sec:loop}, we argue that TC can accommodate a TC Higgs with a physical mass of 125 GeV, with or without effects from walking.
\subsection{Geometric Scaling of the TC Higgs mass}
We will consider at most two-index representations for TC matter,  since at large $N_{\rm TC}$  even higher representations loose quickly asymptotic freedom \cite{Dietrich:2006cm}. The relevant scaling rules are:
\begin{eqnarray}
F_\Pi^2 \sim d(R_{\rm TC})\ m_{\rm TC}^2 \ , \quad v^2=N_{\rm TD}\ F_\Pi^2 \ ,
\label{scaling}
\end{eqnarray}
where $F_\Pi$ is the technipion decay constant, $m_{\rm TC}$ is the dynamically generated constituent techniquark mass, and $v=246$ GeV is the electroweak vacuum expectation value, which will be kept fix in the following.

The squared mass of any large $N_{\rm TC}$ leading technimeson  scales like:
\begin{eqnarray}
(M_H^{TC})^2=   \frac{3}{d(R_{\rm TC})}\ \frac{1}{N_{\rm TD}}\ \frac{v^2}{f_\pi^2}\ m_\sigma^2 \ .
\label{scalingvfixed}
\end{eqnarray}
The leading states for the fundamental (F) representation are fermion-antifermion pairs and in the case of the two-index representations are mesons containing any number of fermions \cite{Sannino:2007yp, Sannino:2008ha}. The normalization to three colors QCD can also be assumed for the two-index representations since the two-index antisymmetric (A) for three colors is exactly QCD and the two-index symmetric (S) at infinite number of TC colors cannot be distinguished from the antisymmetric one.

The  mesonic states which are not leading at large $N_{\rm TC}$ for the fundamental representation will decouple from the leading ones. Their scaling is
 \begin{eqnarray}
(M_H^{TC})^2= \left[\frac{N_{\rm TC}}{3}\right]^p \frac{3}{N_{\rm TC}}\ \frac{1}{N_{\rm TD}}\ \frac{v^2}{f_\pi^2}\ m_\sigma^2 \ ,
\label{scalingvfixedp}
\end{eqnarray}
with $p > 0$ \cite{Sannino:2007yp, Sannino:2008ha}. Applying these general results to the F and the A representation gives the $M_H^{TC}$ scaling formulae summarized in Tab.~\ref{tab:scaling}.
\begin{table}
\begin{center}
\begin{tabular}{c|c|c}
$SU(N_{\rm TC})$ Rep. & $d(R_{\rm TC})$ & $N_{\rm TD}\ (M_H^{TC})^2$ \\
\ & \ \\
\hline
\ & \ \\
$\tiny{\yng(1)}$ & $N_{\rm TC}$ &  $\displaystyle{\frac{v^2}{f_\pi^2}
\left[\frac{3}{N_{\rm TC}}\right]^{1-p}\ m_\sigma^2}$ \\
\ & \ \\
\hline
\ & \ \\
$\tiny{\yng(1,1)}$ & $\displaystyle{\frac{N_{\rm TC}(N_{\rm TC}-1)}{2}}$ &
$\displaystyle{\frac{v^2}{f_\pi^2}\ \frac{3}{N_{\rm TC}(N_{\rm TC}-1)/2}\ m_\sigma^2}$ \\
\ & \ \\
\hline
\ & \ \\
$\tiny{\yng(2)}$ & $\displaystyle{\frac{N_{\rm TC}(N_{\rm TC}+1)}{2}}$ &
$\displaystyle{\frac{v^2}{f_\pi^2}\ \frac{3}{N_{\rm TC}(N_{\rm TC} + 1)/2}\ m_\sigma^2}$ \\
\end{tabular}
\caption{Scaling formulae for $M_H^{TC}$ obtained by scaling up the mass of the QCD $\sigma$ meson. The general formula is given in (\ref{scalingvfixed}), and here is applied to the F representation, with $p=0$, the A representation, and  the S representation.}
\label{tab:scaling}
\end{center}
\end{table}

The geometric scaling above can be compared to the dynamical mass of the TC Higgs required to fit the experiments once the electroweak corrections have been subtracted. Using (\ref{quadcorMrho})  gives \begin{equation}
N_{\rm TD}\ (M_H^{TC})^2 = N_{\rm TD}\ M_H^2 +12\ \kappa^2 r_t^2 m_t^2-6\ \kappa^2 s_\pi\left(m_W^2+\frac{m_Z^2}{2}\right) \ .
\label{eq:NTDMH}
\end{equation}
 The top contribution dominates for ${\cal O}(1)$ values of $N_{\rm TD}$ and $\kappa r_t$. With this assumption the contribution of the gauge bosons can be neglected.
Using (\ref{scalingvfixed})  we estimate $\kappa r_t$, up to corrections due to the weak gauge boson exchange, to be
\begin{equation}
\kappa^2 r_t^2 = \frac{A}{B} \frac{1}{d(R_{\rm TC})} - \frac{N_{\rm TD}}{B}   \qquad {\rm with} \qquad
 A = 3\frac{v^2}{M^2_H} \frac{m^2_{\sigma}}{ f^2_{\pi}} \ , \qquad B = 12 \frac{m^2_t}{M^2_H} \ .
 \label{krt}
\end{equation}
For the case in which $\kappa $ is taken to scale like $\hat{\kappa}/\sqrt{d(R_{\rm TC})}$ we get
\begin{equation}
\hat{\kappa}^2 r_t^2 = \frac{A}{B} - \frac{N_{\rm TD} \, d(R_{\rm TC})}{B}   \ .
\label{hkrt}
\end{equation}
For example, choosing $N_{\rm TD} =1$ and  the S representation with $N_{TC} = 3 $, we have that $d(R_{\rm TC}) = 6$. Using (\ref{krt},\ref{hkrt}) this gives $\kappa r_{t} \simeq 1.5 $ ($\hat{\kappa} r_{t} \simeq 3.8$).

\begin{figure}[t!]
\par
\begin{center}
\includegraphics[width=12.0cm]{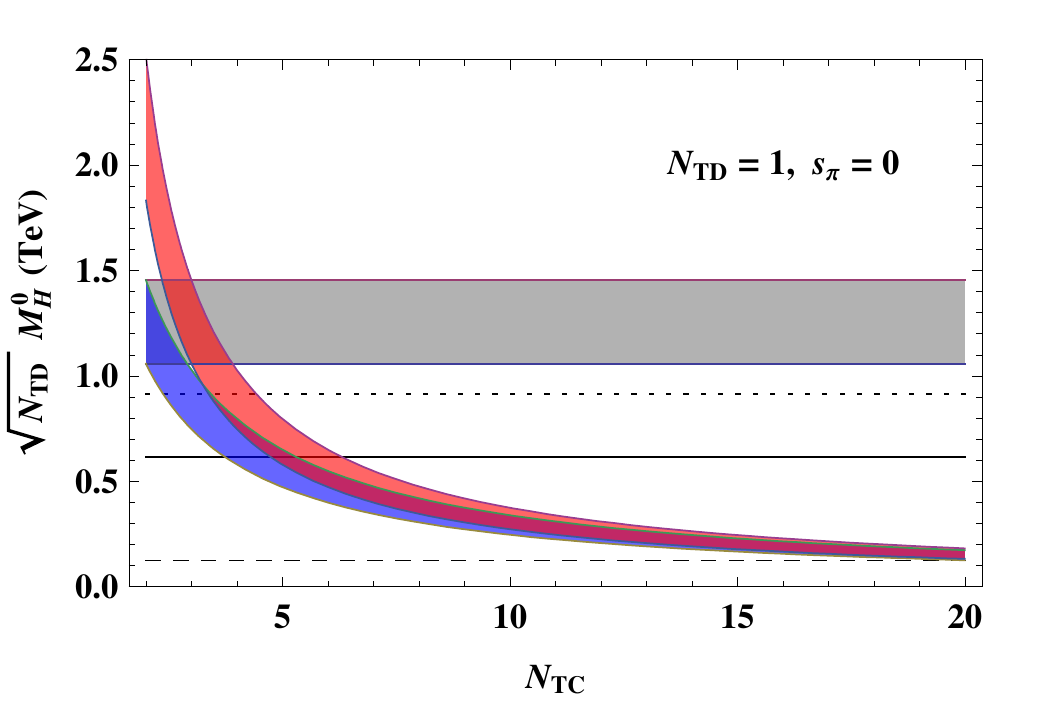}
\end{center}
\caption{Estimates of the dynamical TC Higgs mass $M_H^{TC}$. The bands are constructed using the geometric scaling of Tab.~\ref{tab:scaling}, with gray, blue, and red band for F, A, and S representation, respectively. In each band the lower (upper) curve corresponds to the experimental lower (upper) bound on $m_\sigma$. The dashed, solid, and dotted curves show the required dynamical TC Higgs mass to achieve a 125 GeV Higgs, using (\ref{eq:NTDMH}) for $N_{\rm TD}=1$, $s_\pi=0$, and $\kappa r_t=0$, $\kappa r_t=1$, and $\kappa r_t=1.5  $, respectively.  }
\label{Eq:Ms}
\end{figure}

In figure~\ref{Eq:Ms} we plot the required dynamical mass $\sqrt{N_{\rm TD}} M_H^{TC}$ to fit the observed resonance value for different values of $\kappa r_{t}$. Since the dependence on $N_{\rm TD}$ is small we will show only the case of $N_{\rm TD} =1$ and neglect, by setting $s_{\pi} = 0$, the weak gauge boson corrections. The horizontal lines correspond to $\kappa r_t=0$ (dashed), $\kappa r_t=1 $ (solid), and $\kappa  r_t=1.5 $ (dotted). The estimates of Tab.~\ref{tab:scaling} are shown as the colored bands in the same figure, with the lower (upper) edges of the bands corresponding to the experimental lower (upper) bound on the QCD $m_\sigma$. The horizontal band is for the F representation and $p=1$, the blue (middle) band for the A representation, and the red (lower) band for the S representation.
From Fig.~\ref{Eq:Ms} we observe that if we {\em do} take the external radiative corrections into account, and set $\kappa r_t=1$ ($\kappa r_t=1.5$), then the geometric TC scaling can accommodate a 125 GeV Higgs. This occurs where the bands overlap with the middle solid (upper dotted) horizontal line. This requires $4\leq N_{\rm TC}\leq 5$ ($3\leq N_{\rm TC}\leq 4$) for the S representation, and $5\leq N_{\rm TC}\leq 6$ ($4\leq N_{\rm TC}\leq 5$) for the A representation.  On the other hand, If we do not take the external radiative corrections into account, a TC theory with geometric scaling can only accommodate a 125 GeV Higgs where the bands overlap with the lower dashed horizontal line, requiring large values of $N_{\rm TC}\gtrsim 20$ for the S and A representation.

We conclude that radiative SM corrections cannot be neglected when discussing extensions of the SM featuring TC Higgs states, or any other extension similar to TC. Furthermore these corrections tend to reduce the physical mass allowing for more natural values of the underlying dynamical mass of the TC isosinglet scalar.

\subsection{Walking effects on the TC Higgs mass}\label{dynamics}

In models with walking dynamics the above scalings are expected to overestimate the mass of the TC Higgs. In fact, as $N_{\rm TC}$ and/or $N_{\rm TD}$ take on different values, gauge dynamics is expected to change. For a given TC gauge group and a given representation for the technifermions, there exists a critical number of Dirac techniflavors, $N_{\rm TF}^c$, such that for $N_{\rm TF}<N_{\rm TF}^c$ chiral symmetry is broken, whereas for $N_{\rm TF}>N_{\rm TF}^c$ chiral symmetry is restored. For $N_{\rm TF}$ slightly above $N_{\rm TF}^c$, the gauge theory develops an infrared fixed point. Gauge dynamics becomes conformal at low energies, and the coupling never reaches the critical value for chiral symmetry breaking. For $N_{\rm TF}$ slightly below $N_{\rm TF}^c$, the TC force does break chiral symmetry, but may feel the presence of the nearby fixed point. In this case the theory becomes walking in a range of energies above the chiral symmetry breaking scale. Assuming a continuous phase transition, as $N_{\rm TF}$ approaches $N_{\rm TF}^c$ from below, the technihadron masses and decay constants approach zero
\footnote{There is still the possibility that the phase transition is of {\it jumping} type \cite{Sannino:2012wy,deForcrand:2012vh}, meaning the transition is not continuous and therefore no massive state becomes light when approaching the conformal boundary. Jumping conformal phase transitions have been identified in \cite{Antipin:2012sm}. It is relevant to mention that the first four-dimensional continuous conformal phase transition \'a la Miransky \cite{Miransky:1984ef,Miransky:1988gk,Miransky:1996pd} as function of the number of flavors of the theory has been discovered in \cite{Antipin:2012kc}. For interesting lower dimensional models displaying Miransky scaling investigated on the lattice we refer to \cite{deForcrand:2012se,Nogradi:2012dj}.
}.
It is reasonable to assume that for walking theories when $N_{\rm TF}$ is very close to $N_{\rm TF}^c$ one has \cite{Dietrich:2005jn}
\begin{equation}
(M_H^{TC})^2 \propto (N_{\rm TF}^c-N_{\rm TF})^{\nu_{_H}}
\frac{F_\Pi^2}{d(R_{\rm TC})}   \ ,
\end{equation}
where $\nu_{_H}$ depends on the specific underlying theory. Using the second of (\ref{scaling}) to relate $F_\Pi$ to $v$ one deduces
\begin{equation}
N_{\rm TD}\ (M_H^{TC})^2 \propto (N_{\rm TF}^c-N_{\rm TF})^{\nu_{_H}}  \frac{v^2}{d(R_{\rm TC})} \ .
\label{eq:NCscaling}
\end{equation}
Therefore, if $\nu_{_H}>0$ the Higgs becomes lighter   relative to $v=246$ GeV as $N_{\rm TF}$ gets closer to $N_{\rm TF}^c$.

Different model computations have been used to argue whether the ratio $N_{\rm TD} (M_H^{TC})^2/v^2$ becomes smaller as $N_{\rm TF}\to N_{\rm TF}^c$. These include the technidilaton~\cite{Yamawaki:1985zg,Dietrich:2005jn,Appelquist:2010gy}, truncated Schwinger-Dyson equations~\cite{Gusynin:1989jj,Holdom:1986ub,Holdom:1987yu,Harada:2003dc,Kurachi:2006ej,Doff:2008xx,Doff:2009nk,Doff:2009kq,Doff:2009na}, and computations in orientifold-like theories~\cite{Sannino:2003xe,Hong:2004td}. Perturbative determinations of the conformal window have also shown to lead to a calculable dilaton state parametrically lighter than the other states in the theory \cite{Grinstein:2011dq,Antipin:2011aa,Antipin:2012sm}.

{Albeit these reductions from walking dynamics are welcome, Fig.~\ref{Eq:Ms} shows that one does not need a large suppression of the TC Higgs (dilaton) mass from walking dynamics when the SM radiative corrections are taken into account.}

\subsection{Low energy constraints}
Recent electroweak precision fits of the $S,T,U$ parameters \cite{Kennedy:1988sn,Peskin:1990zt} have been performed in e.g. \cite{Beringer:1900zz,Baak:2012kk}.
In \cite{Beringer:1900zz} the Higgs mass window was chosen to be $115.5 \, {\rm GeV} <m_{h,ref}< 127 \, {\rm GeV}$ with $U=0$ fixed. In \cite{Baak:2012kk}, the Higgs and top masses were fixed at $m_{h,ref}=126$ GeV, $m_{t, ref}=173$ GeV. The resulting fits give

\begin{align}
S&=0.04 \pm 0.09 \ , \quad T=0.07 \pm 0.08  \ , \quad U=0  \quad \mbox{\cite{Beringer:1900zz}} \\
S&=0.03 \pm 0.1 \ , \quad T=0.05 \pm 0.12  \ , \quad U=0.03 \pm 0.10  \quad \mbox{\cite{Baak:2012kk}}
\end{align}
The precision observables are here defined after subtracting the SM contributions.
The $S$ parameter contribution from physics beyond the SM is thus allowed to be as large as $S \lesssim 0.25,(0.35)$ within the 95 (99) $\%$ CL ($S$ and $T$ strongly correlated). 

A difference with respect to the estimates of the precision observables for TC given in e.g. \cite{Peskin:1990zt} is that the reference TC Higgs mass scale is $126$ GeV and not the TeV scale after taking the top corrections into account.  This has the effect of increasing the value of the experimental $S$ parameter with respect to the  earlier estimates for TC and allows for values of the $S$ parameter closer to the ones stemming from minimal models of TC \cite{Sannino:2004qp,Dietrich:2005jn,Dietrich:2005wk,Dietrich:2006cm,Foadi:2007ue,Foadi:2007se,Ryttov:2008xe}. The electroweak precision parameters have been recently studied in more detail in \cite{Foadi:2012ga,Pich:2012dv} using the effective Lagrangian in Eq.~\eqref{eq:L}. 

We have also developed the formalism which makes use of the effective Lagrangian for the TC Higgs allowing to precisely link the intrinsic underlying contribution with the experimentally relevant precision parameters. Flavor changing neutral currents can be minimized by requiring the dynamics to be of walking type~\cite{Holdom:1981rm,Yamawaki:1985zg,Appelquist:1986an}.

\subsection{A candidate model}
As a candidate walking TC model we consider the Next to Minimal Walking Technicolor model (NMWT) \cite{Sannino:2004qp} featuring technifermions transforming according to the sextet representation of $SU(3)_{TC}$ gauge group. For this theory the naive $S$-parameter is $\sim \frac{1}{\pi}$ and the intrinsic TC Higgs mass from geometric scaling of QCD is in the range of $ 0.7$ to $1$~TeV. Therefore, in this theory a physical TC Higgs mass of $126$~ GeV would arise for $\kappa r_t\sim 1$. The corresponding LHC phenomenology has been studied in e.g. \cite{Belyaev:2008yj,Foadi:2008ci,Frandsen:2012rj}. To distinguish these models from a SM Higgs one can use the presence of additional resonances. In particular we expect new spin one resonances coupled to the TC Higgs itself that, e.g., modify the production of the Higgs in association with a SM gauge boson \cite{Belyaev:2008yj,Frandsen:2012rj}. Preliminary lattice results for this model \cite{Fodor:2012ty,Fodor:2012ni} indicate that the masses of the vector  and axial spin one resonances are respectively $M_{\rho} \simeq 1754 \pm 104$~GeV and $M_{A_1} \simeq 2327 \pm 121$~GeV. These values of the spin one masses are within the discovery potential of the LHC \cite{Belyaev:2008yj,Frandsen:2012rj}.

\section{Conclusions}\label{sec:conclusions}

We have argued that the observed Higgs mass at the LHC can be interpreted as  a TC Higgs with a TeV scale dynamical mass. This is so since the SM top-induced radiative corrections reduce the TC Higgs dynamical mass towards the observed value. We used the phenomenologically motivated assumption that the TC Higgs coupling to the top quark is close to the SM value. In this scenario the next non-Goldstone technimesons to be discovered at the LHC have a mass of the order of two to three TeV.

We then investigated the general conditions for the TC spectrum to feature the isosinglet scalar to be identified with the TC Higgs. We reviewed geometric scaling laws in $SU(N_{\rm TC})$ gauge theories, using two-flavor QCD and the $\sigma$ meson as reference. Higher-dimensional representations such as the two-index symmetric and the two-index antisymmetric feature a TC Higgs whose dynamical mass decreases as $N_{\rm TC}$ grows. We showed that this leads to a TC-Higgs dynamical mass compatible with a 125 GeV physical mass already for small values of $N_{\rm TC}$. 

Finally we also reviewed the potential effects on the TC Higgs properties from walking dynamics and argued that they do not need to be large for achieving a phenomenologically viable TC Higgs physical mass once the SM radiative corrections are taken into account. We also discussed a candidate model able to realize the scenario envisioned here with a preliminary prediction for the first vector-resonances based on lattice results. The crucial test of models realizing a light TC Higgs remains the presence of additional resonances whose mass scale cannot be much above $4 \pi v \simeq 3$ TeV --- the highest natural scale for Technicolor.

\subsection*{Acknowledgments}
The work of R.F. is supported by the Marie Curie IIF grant proposal 275012.
The work of M.T.F is supported by a Sapere Aude Grant.

\bibliography{lch}

\begin{thebibliography}{69}%
\makeatletter
\providecommand \@ifxundefined [1]{%
 \@ifx{#1\undefined}
}%
\providecommand \@ifnum [1]{%
 \ifnum #1\expandafter \@firstoftwo
 \else \expandafter \@secondoftwo
 \fi
}%
\providecommand \@ifx [1]{%
 \ifx #1\expandafter \@firstoftwo
 \else \expandafter \@secondoftwo
 \fi
}%
\providecommand \natexlab [1]{#1}%
\providecommand \enquote  [1]{``#1''}%
\providecommand \bibnamefont  [1]{#1}%
\providecommand \bibfnamefont [1]{#1}%
\providecommand \citenamefont [1]{#1}%
\providecommand \href@noop [0]{\@secondoftwo}%
\providecommand \href [0]{\begingroup \@sanitize@url \@href}%
\providecommand \@href[1]{\@@startlink{#1}\@@href}%
\providecommand \@@href[1]{\endgroup#1\@@endlink}%
\providecommand \@sanitize@url [0]{\catcode `\\12\catcode `\$12\catcode
  `\&12\catcode `\#12\catcode `\^12\catcode `\_12\catcode `\%12\relax}%
\providecommand \@@startlink[1]{}%
\providecommand \@@endlink[0]{}%
\providecommand \url  [0]{\begingroup\@sanitize@url \@url }%
\providecommand \@url [1]{\endgroup\@href {#1}{\urlprefix }}%
\providecommand \urlprefix  [0]{URL }%
\providecommand \Eprint [0]{\href }%
\providecommand \doibase [0]{http://dx.doi.org/}%
\providecommand \selectlanguage [0]{\@gobble}%
\providecommand \bibinfo  [0]{\@secondoftwo}%
\providecommand \bibfield  [0]{\@secondoftwo}%
\providecommand \translation [1]{[#1]}%
\providecommand \BibitemOpen [0]{}%
\providecommand \bibitemStop [0]{}%
\providecommand \bibitemNoStop [0]{.\EOS\space}%
\providecommand \EOS [0]{\spacefactor3000\relax}%
\providecommand \BibitemShut  [1]{\csname bibitem#1\endcsname}%
\let\auto@bib@innerbib\@empty
%</preamble>
\bibitem [{\citenamefont {Aad}\ \emph {et~al.}(2012)\citenamefont {Aad} \emph
  {et~al.}}]{:2012gk}%
  \BibitemOpen
  \bibfield  {author} {\bibinfo {author} {\bibfnamefont {G.}~\bibnamefont
  {Aad}} \emph {et~al.} (\bibinfo {collaboration} {ATLAS Collaboration}),\
  }\href@noop {} {\  (\bibinfo {year} {2012})},\ \Eprint
  {http://arxiv.org/abs/1207.7214} {arXiv:1207.7214 [hep-ex]} \BibitemShut
  {NoStop}%
%%CITATION = ARXIV:1207.7214;%%
\bibitem [{\citenamefont {Chatrchyan}\ \emph {et~al.}(2012)\citenamefont
  {Chatrchyan} \emph {et~al.}}]{:2012gu}%
  \BibitemOpen
  \bibfield  {author} {\bibinfo {author} {\bibfnamefont {S.}~\bibnamefont
  {Chatrchyan}} \emph {et~al.} (\bibinfo {collaboration} {CMS Collaboration}),\
  }\href@noop {} {\bibfield  {journal} {\bibinfo  {journal} {Phys.Lett.B}\ }
  (\bibinfo {year} {2012})},\ \Eprint {http://arxiv.org/abs/1207.7235}
  {arXiv:1207.7235 [hep-ex]} \BibitemShut {NoStop}%
%%CITATION = ARXIV:1207.7235;%%
\bibitem [{\citenamefont {Frandsen}\ and\ \citenamefont
  {Sannino}(2012)}]{Frandsen:2012rj}%
  \BibitemOpen
  \bibfield  {author} {\bibinfo {author} {\bibfnamefont {M.~T.}\ \bibnamefont
  {Frandsen}}\ and\ \bibinfo {author} {\bibfnamefont {F.}~\bibnamefont
  {Sannino}},\ }\href@noop {} {\  (\bibinfo {year} {2012})},\ \Eprint
  {http://arxiv.org/abs/1203.3988} {arXiv:1203.3988 [hep-ph]} \BibitemShut
  {NoStop}%
%%CITATION = ARXIV:1203.3988;%%
\bibitem [{\citenamefont {Coleppa}\ \emph {et~al.}(2012)\citenamefont
  {Coleppa}, \citenamefont {Kumar},\ and\ \citenamefont
  {Logan}}]{Coleppa:2012eh}%
  \BibitemOpen
  \bibfield  {author} {\bibinfo {author} {\bibfnamefont {B.}~\bibnamefont
  {Coleppa}}, \bibinfo {author} {\bibfnamefont {K.}~\bibnamefont {Kumar}}, \
  and\ \bibinfo {author} {\bibfnamefont {H.~E.}\ \bibnamefont {Logan}},\
  }\href@noop {} {\  (\bibinfo {year} {2012})},\ \Eprint
  {http://arxiv.org/abs/1208.2692} {arXiv:1208.2692 [hep-ph]} \BibitemShut
  {NoStop}%
%%CITATION = ARXIV:1208.2692;%%
\bibitem [{\citenamefont {Eichten}\ \emph {et~al.}(2012)\citenamefont
  {Eichten}, \citenamefont {Lane},\ and\ \citenamefont
  {Martin}}]{Eichten:2012qb}%
  \BibitemOpen
  \bibfield  {author} {\bibinfo {author} {\bibfnamefont {E.}~\bibnamefont
  {Eichten}}, \bibinfo {author} {\bibfnamefont {K.}~\bibnamefont {Lane}}, \
  and\ \bibinfo {author} {\bibfnamefont {A.}~\bibnamefont {Martin}},\
  }\href@noop {} {\  (\bibinfo {year} {2012})},\ \Eprint
  {http://arxiv.org/abs/1210.5462} {arXiv:1210.5462 [hep-ph]} \BibitemShut
  {NoStop}%
%%CITATION = ARXIV:1210.5462;%%
\bibitem [{\citenamefont {Weinberg}(1976)}]{Weinberg:1975gm}%
  \BibitemOpen
  \bibfield  {author} {\bibinfo {author} {\bibfnamefont {S.}~\bibnamefont
  {Weinberg}},\ }\href {\doibase 10.1103/PhysRevD.13.974} {\bibfield  {journal}
  {\bibinfo  {journal} {Phys.Rev.}\ }\textbf {\bibinfo {volume} {D13}},\
  \bibinfo {pages} {974} (\bibinfo {year} {1976})}\BibitemShut {NoStop}%
%%CITATION = PHRVA,D13,974;%%
\bibitem [{\citenamefont {Dimopoulos}\ and\ \citenamefont
  {Susskind}(1979)}]{Dimopoulos:1979es}%
  \BibitemOpen
  \bibfield  {author} {\bibinfo {author} {\bibfnamefont {S.}~\bibnamefont
  {Dimopoulos}}\ and\ \bibinfo {author} {\bibfnamefont {L.}~\bibnamefont
  {Susskind}},\ }\href {\doibase 10.1016/0550-3213(79)90364-X} {\bibfield
  {journal} {\bibinfo  {journal} {Nucl.Phys.}\ }\textbf {\bibinfo {volume}
  {B155}},\ \bibinfo {pages} {237} (\bibinfo {year} {1979})}\BibitemShut
  {NoStop}%
%%CITATION = NUPHA,B155,237;%%
\bibitem [{\citenamefont {Eichten}\ and\ \citenamefont
  {Lane}(1980)}]{Eichten:1979ah}%
  \BibitemOpen
  \bibfield  {author} {\bibinfo {author} {\bibfnamefont {E.}~\bibnamefont
  {Eichten}}\ and\ \bibinfo {author} {\bibfnamefont {K.~D.}\ \bibnamefont
  {Lane}},\ }\href {\doibase 10.1016/0370-2693(80)90065-9} {\bibfield
  {journal} {\bibinfo  {journal} {Phys.Lett.}\ }\textbf {\bibinfo {volume}
  {B90}},\ \bibinfo {pages} {125} (\bibinfo {year} {1980})}\BibitemShut
  {NoStop}%
%%CITATION = PHLTA,B90,125;%%
\bibitem [{\citenamefont {Sannino}(2008)}]{Sannino:2008ha}%
  \BibitemOpen
  \bibfield  {author} {\bibinfo {author} {\bibfnamefont {F.}~\bibnamefont
  {Sannino}},\ }\href@noop {} {\  (\bibinfo {year} {2008})},\ \Eprint
  {http://arxiv.org/abs/0804.0182} {arXiv:0804.0182 [hep-ph]} \BibitemShut
  {NoStop}%
%%CITATION = ARXIV:0804.0182;%%
\bibitem [{\citenamefont {Sannino}(2009{\natexlab{a}})}]{Sannino:2009za}%
  \BibitemOpen
  \bibfield  {author} {\bibinfo {author} {\bibfnamefont {F.}~\bibnamefont
  {Sannino}},\ }\href@noop {} {\bibfield  {journal} {\bibinfo  {journal} {Acta
  Phys.Polon.}\ }\textbf {\bibinfo {volume} {B40}},\ \bibinfo {pages} {3533}
  (\bibinfo {year} {2009}{\natexlab{a}})},\ \Eprint
  {http://arxiv.org/abs/0911.0931} {arXiv:0911.0931 [hep-ph]} \BibitemShut
  {NoStop}%
%%CITATION = ARXIV:0911.0931;%%
\bibitem [{\citenamefont {Harada}\ \emph {et~al.}(1996)\citenamefont {Harada},
  \citenamefont {Sannino},\ and\ \citenamefont {Schechter}}]{Harada:1995dc}%
  \BibitemOpen
  \bibfield  {author} {\bibinfo {author} {\bibfnamefont {M.}~\bibnamefont
  {Harada}}, \bibinfo {author} {\bibfnamefont {F.}~\bibnamefont {Sannino}}, \
  and\ \bibinfo {author} {\bibfnamefont {J.}~\bibnamefont {Schechter}},\ }\href
  {\doibase 10.1103/PhysRevD.54.1991} {\bibfield  {journal} {\bibinfo
  {journal} {Phys.Rev.}\ }\textbf {\bibinfo {volume} {D54}},\ \bibinfo {pages}
  {1991} (\bibinfo {year} {1996})},\ \Eprint
  {http://arxiv.org/abs/hep-ph/9511335} {arXiv:hep-ph/9511335 [hep-ph]}
  \BibitemShut {NoStop}%
%%CITATION = HEP-PH/9511335;%%
\bibitem [{\citenamefont {Holdom}(1981)}]{Holdom:1981rm}%
  \BibitemOpen
  \bibfield  {author} {\bibinfo {author} {\bibfnamefont {B.}~\bibnamefont
  {Holdom}},\ }\href {\doibase 10.1103/PhysRevD.24.1441} {\bibfield  {journal}
  {\bibinfo  {journal} {Phys.Rev.}\ }\textbf {\bibinfo {volume} {D24}},\
  \bibinfo {pages} {1441} (\bibinfo {year} {1981})}\BibitemShut {NoStop}%
%%CITATION = PHRVA,D24,1441;%%
\bibitem [{\citenamefont {Appelquist}\ and\ \citenamefont
  {Sannino}(1999)}]{Appelquist:1998xf}%
  \BibitemOpen
  \bibfield  {author} {\bibinfo {author} {\bibfnamefont {T.}~\bibnamefont
  {Appelquist}}\ and\ \bibinfo {author} {\bibfnamefont {F.}~\bibnamefont
  {Sannino}},\ }\href {\doibase 10.1103/PhysRevD.59.067702} {\bibfield
  {journal} {\bibinfo  {journal} {Phys.Rev.}\ }\textbf {\bibinfo {volume}
  {D59}},\ \bibinfo {pages} {067702} (\bibinfo {year} {1999})},\ \Eprint
  {http://arxiv.org/abs/hep-ph/9806409} {arXiv:hep-ph/9806409 [hep-ph]}
  \BibitemShut {NoStop}%
%%CITATION = HEP-PH/9806409;%%
\bibitem [{\citenamefont {Kurachi}\ and\ \citenamefont
  {Shrock}(2006)}]{Kurachi:2006ej}%
  \BibitemOpen
  \bibfield  {author} {\bibinfo {author} {\bibfnamefont {M.}~\bibnamefont
  {Kurachi}}\ and\ \bibinfo {author} {\bibfnamefont {R.}~\bibnamefont
  {Shrock}},\ }\href {\doibase 10.1088/1126-6708/2006/12/034} {\bibfield
  {journal} {\bibinfo  {journal} {JHEP}\ }\textbf {\bibinfo {volume} {0612}},\
  \bibinfo {pages} {034} (\bibinfo {year} {2006})},\ \Eprint
  {http://arxiv.org/abs/hep-ph/0605290} {arXiv:hep-ph/0605290 [hep-ph]}
  \BibitemShut {NoStop}%
%%CITATION = HEP-PH/0605290;%%
\bibitem [{\citenamefont {Sannino}(2010{\natexlab{a}})}]{Sannino:2010ca}%
  \BibitemOpen
  \bibfield  {author} {\bibinfo {author} {\bibfnamefont {F.}~\bibnamefont
  {Sannino}},\ }\href {\doibase 10.1103/PhysRevD.82.081701} {\bibfield
  {journal} {\bibinfo  {journal} {Phys.Rev.}\ }\textbf {\bibinfo {volume}
  {D82}},\ \bibinfo {pages} {081701} (\bibinfo {year} {2010}{\natexlab{a}})},\
  \Eprint {http://arxiv.org/abs/1006.0207} {arXiv:1006.0207 [hep-lat]}
  \BibitemShut {NoStop}%
%%CITATION = ARXIV:1006.0207;%%
\bibitem [{\citenamefont {Sannino}(2010{\natexlab{b}})}]{Sannino:2010fh}%
  \BibitemOpen
  \bibfield  {author} {\bibinfo {author} {\bibfnamefont {F.}~\bibnamefont
  {Sannino}},\ }\href {\doibase 10.1103/PhysRevLett.105.232002} {\bibfield
  {journal} {\bibinfo  {journal} {Phys.Rev.Lett.}\ }\textbf {\bibinfo {volume}
  {105}},\ \bibinfo {pages} {232002} (\bibinfo {year} {2010}{\natexlab{b}})},\
  \Eprint {http://arxiv.org/abs/1007.0254} {arXiv:1007.0254 [hep-ph]}
  \BibitemShut {NoStop}%
%%CITATION = ARXIV:1007.0254;%%
\bibitem [{\citenamefont {Di~Chiara}\ \emph {et~al.}(2011)\citenamefont
  {Di~Chiara}, \citenamefont {Pica},\ and\ \citenamefont
  {Sannino}}]{DiChiara:2010xb}%
  \BibitemOpen
  \bibfield  {author} {\bibinfo {author} {\bibfnamefont {S.}~\bibnamefont
  {Di~Chiara}}, \bibinfo {author} {\bibfnamefont {C.}~\bibnamefont {Pica}}, \
  and\ \bibinfo {author} {\bibfnamefont {F.}~\bibnamefont {Sannino}},\ }\href
  {\doibase 10.1016/j.physletb.2011.05.008} {\bibfield  {journal} {\bibinfo
  {journal} {Phys.Lett.}\ }\textbf {\bibinfo {volume} {B700}},\ \bibinfo
  {pages} {229} (\bibinfo {year} {2011})},\ \Eprint
  {http://arxiv.org/abs/1008.1267} {arXiv:1008.1267 [hep-ph]} \BibitemShut
  {NoStop}%
%%CITATION = ARXIV:1008.1267;%%
\bibitem [{\citenamefont {Foadi}\ and\ \citenamefont
  {Sannino}(2012)}]{Foadi:2012ga}%
  \BibitemOpen
  \bibfield  {author} {\bibinfo {author} {\bibfnamefont {R.}~\bibnamefont
  {Foadi}}\ and\ \bibinfo {author} {\bibfnamefont {F.}~\bibnamefont
  {Sannino}},\ }\href@noop {} {\  (\bibinfo {year} {2012})},\ \Eprint
  {http://arxiv.org/abs/1207.1541} {arXiv:1207.1541 [hep-ph]} \BibitemShut
  {NoStop}%
%%CITATION = ARXIV:1207.1541;%%
\bibitem [{\citenamefont {Yamawaki}\ \emph {et~al.}(1986)\citenamefont
  {Yamawaki}, \citenamefont {Bando},\ and\ \citenamefont
  {Matumoto}}]{Yamawaki:1985zg}%
  \BibitemOpen
  \bibfield  {author} {\bibinfo {author} {\bibfnamefont {K.}~\bibnamefont
  {Yamawaki}}, \bibinfo {author} {\bibfnamefont {M.}~\bibnamefont {Bando}}, \
  and\ \bibinfo {author} {\bibfnamefont {K.-i.}\ \bibnamefont {Matumoto}},\
  }\href {\doibase 10.1103/PhysRevLett.56.1335} {\bibfield  {journal} {\bibinfo
   {journal} {Phys.Rev.Lett.}\ }\textbf {\bibinfo {volume} {56}},\ \bibinfo
  {pages} {1335} (\bibinfo {year} {1986})}\BibitemShut {NoStop}%
%%CITATION = PRLTA,56,1335;%%
\bibitem [{\citenamefont {Bando}\ \emph {et~al.}(1986)\citenamefont {Bando},
  \citenamefont {Matumoto},\ and\ \citenamefont {Yamawaki}}]{Bando:1986bg}%
  \BibitemOpen
  \bibfield  {author} {\bibinfo {author} {\bibfnamefont {M.}~\bibnamefont
  {Bando}}, \bibinfo {author} {\bibfnamefont {K.-i.}\ \bibnamefont {Matumoto}},
  \ and\ \bibinfo {author} {\bibfnamefont {K.}~\bibnamefont {Yamawaki}},\
  }\href {\doibase 10.1016/0370-2693(86)91516-9} {\bibfield  {journal}
  {\bibinfo  {journal} {Phys.Lett.}\ }\textbf {\bibinfo {volume} {B178}},\
  \bibinfo {pages} {308} (\bibinfo {year} {1986})}\BibitemShut {NoStop}%
%%CITATION = PHLTA,B178,308;%%
\bibitem [{\citenamefont {Dietrich}\ \emph {et~al.}(2005)\citenamefont
  {Dietrich}, \citenamefont {Sannino},\ and\ \citenamefont
  {Tuominen}}]{Dietrich:2005jn}%
  \BibitemOpen
  \bibfield  {author} {\bibinfo {author} {\bibfnamefont {D.~D.}\ \bibnamefont
  {Dietrich}}, \bibinfo {author} {\bibfnamefont {F.}~\bibnamefont {Sannino}}, \
  and\ \bibinfo {author} {\bibfnamefont {K.}~\bibnamefont {Tuominen}},\ }\href
  {\doibase 10.1103/PhysRevD.72.055001} {\bibfield  {journal} {\bibinfo
  {journal} {Phys.Rev.}\ }\textbf {\bibinfo {volume} {D72}},\ \bibinfo {pages}
  {055001} (\bibinfo {year} {2005})},\ \Eprint
  {http://arxiv.org/abs/hep-ph/0505059} {arXiv:hep-ph/0505059 [hep-ph]}
  \BibitemShut {NoStop}%
%%CITATION = HEP-PH/0505059;%%
\bibitem [{\citenamefont {Appelquist}\ and\ \citenamefont
  {Bai}(2010)}]{Appelquist:2010gy}%
  \BibitemOpen
  \bibfield  {author} {\bibinfo {author} {\bibfnamefont {T.}~\bibnamefont
  {Appelquist}}\ and\ \bibinfo {author} {\bibfnamefont {Y.}~\bibnamefont
  {Bai}},\ }\href {\doibase 10.1103/PhysRevD.82.071701} {\bibfield  {journal}
  {\bibinfo  {journal} {Phys.Rev.}\ }\textbf {\bibinfo {volume} {D82}},\
  \bibinfo {pages} {071701} (\bibinfo {year} {2010})},\ \Eprint
  {http://arxiv.org/abs/1006.4375} {arXiv:1006.4375 [hep-ph]} \BibitemShut
  {NoStop}%
%%CITATION = ARXIV:1006.4375;%%
\bibitem [{\citenamefont {Belyaev}\ \emph {et~al.}(2009)\citenamefont
  {Belyaev}, \citenamefont {Foadi}, \citenamefont {Frandsen}, \citenamefont
  {Jarvinen}, \citenamefont {Sannino} \emph {et~al.}}]{Belyaev:2008yj}%
  \BibitemOpen
  \bibfield  {author} {\bibinfo {author} {\bibfnamefont {A.}~\bibnamefont
  {Belyaev}}, \bibinfo {author} {\bibfnamefont {R.}~\bibnamefont {Foadi}},
  \bibinfo {author} {\bibfnamefont {M.~T.}\ \bibnamefont {Frandsen}}, \bibinfo
  {author} {\bibfnamefont {M.}~\bibnamefont {Jarvinen}}, \bibinfo {author}
  {\bibfnamefont {F.}~\bibnamefont {Sannino}},  \emph {et~al.},\ }\href
  {\doibase 10.1103/PhysRevD.79.035006} {\bibfield  {journal} {\bibinfo
  {journal} {Phys.Rev.}\ }\textbf {\bibinfo {volume} {D79}},\ \bibinfo {pages}
  {035006} (\bibinfo {year} {2009})},\ \Eprint {http://arxiv.org/abs/0809.0793}
  {arXiv:0809.0793 [hep-ph]} \BibitemShut {NoStop}%
%%CITATION = ARXIV:0809.0793;%%
\bibitem [{\citenamefont {Foadi}\ \emph {et~al.}(2009)\citenamefont {Foadi},
  \citenamefont {Jarvinen},\ and\ \citenamefont {Sannino}}]{Foadi:2008xj}%
  \BibitemOpen
  \bibfield  {author} {\bibinfo {author} {\bibfnamefont {R.}~\bibnamefont
  {Foadi}}, \bibinfo {author} {\bibfnamefont {M.}~\bibnamefont {Jarvinen}}, \
  and\ \bibinfo {author} {\bibfnamefont {F.}~\bibnamefont {Sannino}},\ }\href
  {\doibase 10.1103/PhysRevD.79.035010} {\bibfield  {journal} {\bibinfo
  {journal} {Phys.Rev.}\ }\textbf {\bibinfo {volume} {D79}},\ \bibinfo {pages}
  {035010} (\bibinfo {year} {2009})},\ \Eprint {http://arxiv.org/abs/0811.3719}
  {arXiv:0811.3719 [hep-ph]} \BibitemShut {NoStop}%
%%CITATION = ARXIV:0811.3719;%%
\bibitem [{\citenamefont {Samuel}(1990)}]{Samuel:1990dq}%
  \BibitemOpen
  \bibfield  {author} {\bibinfo {author} {\bibfnamefont {S.}~\bibnamefont
  {Samuel}},\ }\href {\doibase 10.1016/0550-3213(90)90378-Q} {\bibfield
  {journal} {\bibinfo  {journal} {Nucl.Phys.}\ }\textbf {\bibinfo {volume}
  {B347}},\ \bibinfo {pages} {625} (\bibinfo {year} {1990})}\BibitemShut
  {NoStop}%
%%CITATION = NUPHA,B347,625;%%
\bibitem [{\citenamefont {Hill}(1995)}]{Hill:1994hp}%
  \BibitemOpen
  \bibfield  {author} {\bibinfo {author} {\bibfnamefont {C.~T.}\ \bibnamefont
  {Hill}},\ }\href {\doibase 10.1016/0370-2693(94)01660-5} {\bibfield
  {journal} {\bibinfo  {journal} {Phys.Lett.}\ }\textbf {\bibinfo {volume}
  {B345}},\ \bibinfo {pages} {483} (\bibinfo {year} {1995})},\ \Eprint
  {http://arxiv.org/abs/hep-ph/9411426} {arXiv:hep-ph/9411426 [hep-ph]}
  \BibitemShut {NoStop}%
%%CITATION = HEP-PH/9411426;%%
\bibitem [{\citenamefont {Appelquist}\ \emph
  {et~al.}(2004{\natexlab{a}})\citenamefont {Appelquist}, \citenamefont
  {Piai},\ and\ \citenamefont {Shrock}}]{Appelquist:2003hn}%
  \BibitemOpen
  \bibfield  {author} {\bibinfo {author} {\bibfnamefont {T.}~\bibnamefont
  {Appelquist}}, \bibinfo {author} {\bibfnamefont {M.}~\bibnamefont {Piai}}, \
  and\ \bibinfo {author} {\bibfnamefont {R.}~\bibnamefont {Shrock}},\ }\href
  {\doibase 10.1103/PhysRevD.69.015002} {\bibfield  {journal} {\bibinfo
  {journal} {Phys.Rev.}\ }\textbf {\bibinfo {volume} {D69}},\ \bibinfo {pages}
  {015002} (\bibinfo {year} {2004}{\natexlab{a}})},\ \Eprint
  {http://arxiv.org/abs/hep-ph/0308061} {arXiv:hep-ph/0308061 [hep-ph]}
  \BibitemShut {NoStop}%
%%CITATION = HEP-PH/0308061;%%
\bibitem [{\citenamefont {Appelquist}\ \emph
  {et~al.}(2004{\natexlab{b}})\citenamefont {Appelquist}, \citenamefont
  {Christensen}, \citenamefont {Piai},\ and\ \citenamefont
  {Shrock}}]{Appelquist:2004ai}%
  \BibitemOpen
  \bibfield  {author} {\bibinfo {author} {\bibfnamefont {T.}~\bibnamefont
  {Appelquist}}, \bibinfo {author} {\bibfnamefont {N.~D.}\ \bibnamefont
  {Christensen}}, \bibinfo {author} {\bibfnamefont {M.}~\bibnamefont {Piai}}, \
  and\ \bibinfo {author} {\bibfnamefont {R.}~\bibnamefont {Shrock}},\ }\href
  {\doibase 10.1103/PhysRevD.70.093010} {\bibfield  {journal} {\bibinfo
  {journal} {Phys.Rev.}\ }\textbf {\bibinfo {volume} {D70}},\ \bibinfo {pages}
  {093010} (\bibinfo {year} {2004}{\natexlab{b}})},\ \Eprint
  {http://arxiv.org/abs/hep-ph/0409035} {arXiv:hep-ph/0409035 [hep-ph]}
  \BibitemShut {NoStop}%
%%CITATION = HEP-PH/0409035;%%
\bibitem [{\citenamefont {Hashimoto}(2011)}]{Hashimoto:2011ma}%
  \BibitemOpen
  \bibfield  {author} {\bibinfo {author} {\bibfnamefont {M.}~\bibnamefont
  {Hashimoto}},\ }\href {\doibase 10.1103/PhysRevD.83.096003} {\bibfield
  {journal} {\bibinfo  {journal} {Phys.Rev.}\ }\textbf {\bibinfo {volume}
  {D83}},\ \bibinfo {pages} {096003} (\bibinfo {year} {2011})},\ \Eprint
  {http://arxiv.org/abs/1103.5576} {arXiv:1103.5576 [hep-ph]} \BibitemShut
  {NoStop}%
%%CITATION = ARXIV:1103.5576;%%
\bibitem [{\citenamefont {Matsuzaki}\ and\ \citenamefont
  {Yamawaki}(2012)}]{Matsuzaki:2012vc}%
  \BibitemOpen
  \bibfield  {author} {\bibinfo {author} {\bibfnamefont {S.}~\bibnamefont
  {Matsuzaki}}\ and\ \bibinfo {author} {\bibfnamefont {K.}~\bibnamefont
  {Yamawaki}},\ }\href {\doibase 10.1103/PhysRevD.86.035025} {\bibfield
  {journal} {\bibinfo  {journal} {Phys.Rev.}\ }\textbf {\bibinfo {volume}
  {D86}},\ \bibinfo {pages} {035025} (\bibinfo {year} {2012})},\ \Eprint
  {http://arxiv.org/abs/1206.6703} {arXiv:1206.6703 [hep-ph]} \BibitemShut
  {NoStop}%
%%CITATION = ARXIV:1206.6703;%%
\bibitem [{\citenamefont {Beringer}\ \emph {et~al.}(2012)\citenamefont
  {Beringer} \emph {et~al.}}]{Beringer:1900zz}%
  \BibitemOpen
  \bibfield  {author} {\bibinfo {author} {\bibfnamefont {J.}~\bibnamefont
  {Beringer}} \emph {et~al.} (\bibinfo {collaboration} {Particle Data Group}),\
  }\href {\doibase 10.1103/PhysRevD.86.010001} {\bibfield  {journal} {\bibinfo
  {journal} {Phys.Rev.}\ }\textbf {\bibinfo {volume} {D86}},\ \bibinfo {pages}
  {010001} (\bibinfo {year} {2012})}\BibitemShut {NoStop}%
%%CITATION = PHRVA,D86,010001;%%
\bibitem [{\citenamefont {Sannino}(2009{\natexlab{b}})}]{Sannino:2009aw}%
  \BibitemOpen
  \bibfield  {author} {\bibinfo {author} {\bibfnamefont {F.}~\bibnamefont
  {Sannino}},\ }\href {\doibase 10.1103/PhysRevD.79.096007} {\bibfield
  {journal} {\bibinfo  {journal} {Phys.Rev.}\ }\textbf {\bibinfo {volume}
  {D79}},\ \bibinfo {pages} {096007} (\bibinfo {year} {2009}{\natexlab{b}})},\
  \Eprint {http://arxiv.org/abs/0902.3494} {arXiv:0902.3494 [hep-ph]}
  \BibitemShut {NoStop}%
%%CITATION = ARXIV:0902.3494;%%
\bibitem [{\citenamefont {Mojaza}\ \emph {et~al.}(2012)\citenamefont {Mojaza},
  \citenamefont {Pica}, \citenamefont {Ryttov},\ and\ \citenamefont
  {Sannino}}]{Mojaza:2012zd}%
  \BibitemOpen
  \bibfield  {author} {\bibinfo {author} {\bibfnamefont {M.}~\bibnamefont
  {Mojaza}}, \bibinfo {author} {\bibfnamefont {C.}~\bibnamefont {Pica}},
  \bibinfo {author} {\bibfnamefont {T.~A.}\ \bibnamefont {Ryttov}}, \ and\
  \bibinfo {author} {\bibfnamefont {F.}~\bibnamefont {Sannino}},\ }\href@noop
  {} {\  (\bibinfo {year} {2012})},\ \Eprint {http://arxiv.org/abs/1206.2652}
  {arXiv:1206.2652 [hep-ph]} \BibitemShut {NoStop}%
%%CITATION = ARXIV:1206.2652;%%
\bibitem [{\citenamefont {Dietrich}\ and\ \citenamefont
  {Sannino}(2007)}]{Dietrich:2006cm}%
  \BibitemOpen
  \bibfield  {author} {\bibinfo {author} {\bibfnamefont {D.~D.}\ \bibnamefont
  {Dietrich}}\ and\ \bibinfo {author} {\bibfnamefont {F.}~\bibnamefont
  {Sannino}},\ }\href {\doibase 10.1103/PhysRevD.75.085018} {\bibfield
  {journal} {\bibinfo  {journal} {Phys.Rev.}\ }\textbf {\bibinfo {volume}
  {D75}},\ \bibinfo {pages} {085018} (\bibinfo {year} {2007})},\ \Eprint
  {http://arxiv.org/abs/hep-ph/0611341} {arXiv:hep-ph/0611341 [hep-ph]}
  \BibitemShut {NoStop}%
%%CITATION = HEP-PH/0611341;%%
\bibitem [{\citenamefont {Sannino}\ and\ \citenamefont
  {Schechter}(2007)}]{Sannino:2007yp}%
  \BibitemOpen
  \bibfield  {author} {\bibinfo {author} {\bibfnamefont {F.}~\bibnamefont
  {Sannino}}\ and\ \bibinfo {author} {\bibfnamefont {J.}~\bibnamefont
  {Schechter}},\ }\href {\doibase 10.1103/PhysRevD.76.014014} {\bibfield
  {journal} {\bibinfo  {journal} {Phys.Rev.}\ }\textbf {\bibinfo {volume}
  {D76}},\ \bibinfo {pages} {014014} (\bibinfo {year} {2007})},\ \Eprint
  {http://arxiv.org/abs/0704.0602} {arXiv:0704.0602 [hep-ph]} \BibitemShut
  {NoStop}%
%%CITATION = ARXIV:0704.0602;%%
\bibitem [{\citenamefont {Sannino}(2012)}]{Sannino:2012wy}%
  \BibitemOpen
  \bibfield  {author} {\bibinfo {author} {\bibfnamefont {F.}~\bibnamefont
  {Sannino}},\ }\href@noop {} {\  (\bibinfo {year} {2012})},\ \Eprint
  {http://arxiv.org/abs/1205.4246} {arXiv:1205.4246 [hep-ph]} \BibitemShut
  {NoStop}%
%%CITATION = ARXIV:1205.4246;%%
\bibitem [{\citenamefont {de~Forcrand}\ \emph
  {et~al.}(2012{\natexlab{a}})\citenamefont {de~Forcrand}, \citenamefont
  {Kim},\ and\ \citenamefont {Unger}}]{deForcrand:2012vh}%
  \BibitemOpen
  \bibfield  {author} {\bibinfo {author} {\bibfnamefont {P.}~\bibnamefont
  {de~Forcrand}}, \bibinfo {author} {\bibfnamefont {S.}~\bibnamefont {Kim}}, \
  and\ \bibinfo {author} {\bibfnamefont {W.}~\bibnamefont {Unger}},\
  }\href@noop {} {\  (\bibinfo {year} {2012}{\natexlab{a}})},\ \Eprint
  {http://arxiv.org/abs/1208.2148} {arXiv:1208.2148 [hep-lat]} \BibitemShut
  {NoStop}%
%%CITATION = ARXIV:1208.2148;%%
\bibitem [{\citenamefont {Antipin}\ \emph
  {et~al.}(2012{\natexlab{a}})\citenamefont {Antipin}, \citenamefont {Mojaza},\
  and\ \citenamefont {Sannino}}]{Antipin:2012sm}%
  \BibitemOpen
  \bibfield  {author} {\bibinfo {author} {\bibfnamefont {O.}~\bibnamefont
  {Antipin}}, \bibinfo {author} {\bibfnamefont {M.}~\bibnamefont {Mojaza}}, \
  and\ \bibinfo {author} {\bibfnamefont {F.}~\bibnamefont {Sannino}},\
  }\href@noop {} {\  (\bibinfo {year} {2012}{\natexlab{a}})},\ \Eprint
  {http://arxiv.org/abs/1208.0987} {arXiv:1208.0987 [hep-ph]} \BibitemShut
  {NoStop}%
%%CITATION = ARXIV:1208.0987;%%
\bibitem [{\citenamefont {Miransky}(1985)}]{Miransky:1984ef}%
  \BibitemOpen
  \bibfield  {author} {\bibinfo {author} {\bibfnamefont {V.}~\bibnamefont
  {Miransky}},\ }\href {\doibase 10.1007/BF02724229} {\bibfield  {journal}
  {\bibinfo  {journal} {Nuovo Cim.}\ }\textbf {\bibinfo {volume} {A90}},\
  \bibinfo {pages} {149} (\bibinfo {year} {1985})}\BibitemShut {NoStop}%
%%CITATION = NUCIA,A90,149;%%
\bibitem [{\citenamefont {Miransky}\ and\ \citenamefont
  {Yamawaki}(1989)}]{Miransky:1988gk}%
  \BibitemOpen
  \bibfield  {author} {\bibinfo {author} {\bibfnamefont {V.}~\bibnamefont
  {Miransky}}\ and\ \bibinfo {author} {\bibfnamefont {K.}~\bibnamefont
  {Yamawaki}},\ }\href {\doibase 10.1142/S0217732389000186} {\bibfield
  {journal} {\bibinfo  {journal} {Mod.Phys.Lett.}\ }\textbf {\bibinfo {volume}
  {A4}},\ \bibinfo {pages} {129} (\bibinfo {year} {1989})}\BibitemShut
  {NoStop}%
%%CITATION = MPLAE,A4,129;%%
\bibitem [{\citenamefont {Miransky}\ and\ \citenamefont
  {Yamawaki}(1997)}]{Miransky:1996pd}%
  \BibitemOpen
  \bibfield  {author} {\bibinfo {author} {\bibfnamefont {V.}~\bibnamefont
  {Miransky}}\ and\ \bibinfo {author} {\bibfnamefont {K.}~\bibnamefont
  {Yamawaki}},\ }\href {\doibase 10.1103/PhysRevD.56.3768,
  10.1103/PhysRevD.55.5051} {\bibfield  {journal} {\bibinfo  {journal}
  {Phys.Rev.}\ }\textbf {\bibinfo {volume} {D55}},\ \bibinfo {pages} {5051}
  (\bibinfo {year} {1997})},\ \Eprint {http://arxiv.org/abs/hep-th/9611142}
  {arXiv:hep-th/9611142 [hep-th]} \BibitemShut {NoStop}%
%%CITATION = HEP-TH/9611142;%%
\bibitem [{\citenamefont {Antipin}\ \emph
  {et~al.}(2012{\natexlab{b}})\citenamefont {Antipin}, \citenamefont
  {Di~Chiara}, \citenamefont {Mojaza}, \citenamefont {Molgaard},\ and\
  \citenamefont {Sannino}}]{Antipin:2012kc}%
  \BibitemOpen
  \bibfield  {author} {\bibinfo {author} {\bibfnamefont {O.}~\bibnamefont
  {Antipin}}, \bibinfo {author} {\bibfnamefont {S.}~\bibnamefont {Di~Chiara}},
  \bibinfo {author} {\bibfnamefont {M.}~\bibnamefont {Mojaza}}, \bibinfo
  {author} {\bibfnamefont {E.}~\bibnamefont {Molgaard}}, \ and\ \bibinfo
  {author} {\bibfnamefont {F.}~\bibnamefont {Sannino}},\ }\href {\doibase
  10.1103/PhysRevD.86.085009} {\bibfield  {journal} {\bibinfo  {journal}
  {Phys.Rev.}\ }\textbf {\bibinfo {volume} {D86}},\ \bibinfo {pages} {085009}
  (\bibinfo {year} {2012}{\natexlab{b}})},\ \Eprint
  {http://arxiv.org/abs/1205.6157} {arXiv:1205.6157 [hep-ph]} \BibitemShut
  {NoStop}%
%%CITATION = ARXIV:1205.6157;%%
\bibitem [{\citenamefont {de~Forcrand}\ \emph
  {et~al.}(2012{\natexlab{b}})\citenamefont {de~Forcrand}, \citenamefont
  {Pepe},\ and\ \citenamefont {Wiese}}]{deForcrand:2012se}%
  \BibitemOpen
  \bibfield  {author} {\bibinfo {author} {\bibfnamefont {P.}~\bibnamefont
  {de~Forcrand}}, \bibinfo {author} {\bibfnamefont {M.}~\bibnamefont {Pepe}}, \
  and\ \bibinfo {author} {\bibfnamefont {U.}~\bibnamefont {Wiese}},\ }\href
  {\doibase 10.1103/PhysRevD.86.075006} {\bibfield  {journal} {\bibinfo
  {journal} {Phys.Rev.}\ }\textbf {\bibinfo {volume} {D86}},\ \bibinfo {pages}
  {075006} (\bibinfo {year} {2012}{\natexlab{b}})},\ \Eprint
  {http://arxiv.org/abs/1204.4913} {arXiv:1204.4913 [hep-lat]} \BibitemShut
  {NoStop}%
%%CITATION = ARXIV:1204.4913;%%
\bibitem [{\citenamefont {Nogradi}(2012)}]{Nogradi:2012dj}%
  \BibitemOpen
  \bibfield  {author} {\bibinfo {author} {\bibfnamefont {D.}~\bibnamefont
  {Nogradi}},\ }\href {\doibase 10.1007/JHEP05(2012)089} {\bibfield  {journal}
  {\bibinfo  {journal} {JHEP}\ }\textbf {\bibinfo {volume} {1205}},\ \bibinfo
  {pages} {089} (\bibinfo {year} {2012})},\ \Eprint
  {http://arxiv.org/abs/1202.4616} {arXiv:1202.4616 [hep-lat]} \BibitemShut
  {NoStop}%
%%CITATION = ARXIV:1202.4616;%%
\bibitem [{\citenamefont {Gusynin}\ and\ \citenamefont
  {Miransky}(1989)}]{Gusynin:1989jj}%
  \BibitemOpen
  \bibfield  {author} {\bibinfo {author} {\bibfnamefont {V.}~\bibnamefont
  {Gusynin}}\ and\ \bibinfo {author} {\bibfnamefont {V.}~\bibnamefont
  {Miransky}},\ }\href@noop {} {\bibfield  {journal} {\bibinfo  {journal}
  {Sov.Phys.JETP}\ }\textbf {\bibinfo {volume} {68}},\ \bibinfo {pages} {232}
  (\bibinfo {year} {1989})}\BibitemShut {NoStop}%
%%CITATION = SPHJA,68,232;%%
\bibitem [{\citenamefont {Holdom}\ and\ \citenamefont
  {Terning}(1987)}]{Holdom:1986ub}%
  \BibitemOpen
  \bibfield  {author} {\bibinfo {author} {\bibfnamefont {B.}~\bibnamefont
  {Holdom}}\ and\ \bibinfo {author} {\bibfnamefont {J.}~\bibnamefont
  {Terning}},\ }\href {\doibase 10.1016/0370-2693(87)91109-9} {\bibfield
  {journal} {\bibinfo  {journal} {Phys.Lett.}\ }\textbf {\bibinfo {volume}
  {B187}},\ \bibinfo {pages} {357} (\bibinfo {year} {1987})}\BibitemShut
  {NoStop}%
%%CITATION = PHLTA,B187,357;%%
\bibitem [{\citenamefont {Holdom}\ and\ \citenamefont
  {Terning}(1988)}]{Holdom:1987yu}%
  \BibitemOpen
  \bibfield  {author} {\bibinfo {author} {\bibfnamefont {B.}~\bibnamefont
  {Holdom}}\ and\ \bibinfo {author} {\bibfnamefont {J.}~\bibnamefont
  {Terning}},\ }\href {\doibase 10.1016/0370-2693(88)90783-6} {\bibfield
  {journal} {\bibinfo  {journal} {Phys.Lett.}\ }\textbf {\bibinfo {volume}
  {B200}},\ \bibinfo {pages} {338} (\bibinfo {year} {1988})}\BibitemShut
  {NoStop}%
%%CITATION = PHLTA,B200,338;%%
\bibitem [{\citenamefont {Harada}\ \emph {et~al.}(2003)\citenamefont {Harada},
  \citenamefont {Kurachi},\ and\ \citenamefont {Yamawaki}}]{Harada:2003dc}%
  \BibitemOpen
  \bibfield  {author} {\bibinfo {author} {\bibfnamefont {M.}~\bibnamefont
  {Harada}}, \bibinfo {author} {\bibfnamefont {M.}~\bibnamefont {Kurachi}}, \
  and\ \bibinfo {author} {\bibfnamefont {K.}~\bibnamefont {Yamawaki}},\ }\href
  {\doibase 10.1103/PhysRevD.68.076001} {\bibfield  {journal} {\bibinfo
  {journal} {Phys.Rev.}\ }\textbf {\bibinfo {volume} {D68}},\ \bibinfo {pages}
  {076001} (\bibinfo {year} {2003})},\ \Eprint
  {http://arxiv.org/abs/hep-ph/0305018} {arXiv:hep-ph/0305018 [hep-ph]}
  \BibitemShut {NoStop}%
%%CITATION = HEP-PH/0305018;%%
\bibitem [{\citenamefont {Doff}\ \emph {et~al.}(2008)\citenamefont {Doff},
  \citenamefont {Natale},\ and\ \citenamefont {Rodrigues~da
  Silva}}]{Doff:2008xx}%
  \BibitemOpen
  \bibfield  {author} {\bibinfo {author} {\bibfnamefont {A.}~\bibnamefont
  {Doff}}, \bibinfo {author} {\bibfnamefont {A.}~\bibnamefont {Natale}}, \ and\
  \bibinfo {author} {\bibfnamefont {P.}~\bibnamefont {Rodrigues~da Silva}},\
  }\href {\doibase 10.1103/PhysRevD.77.075012} {\bibfield  {journal} {\bibinfo
  {journal} {Phys.Rev.}\ }\textbf {\bibinfo {volume} {D77}},\ \bibinfo {pages}
  {075012} (\bibinfo {year} {2008})},\ \Eprint {http://arxiv.org/abs/0802.1898}
  {arXiv:0802.1898 [hep-ph]} \BibitemShut {NoStop}%
%%CITATION = ARXIV:0802.1898;%%
\bibitem [{\citenamefont {Doff}\ and\ \citenamefont
  {Natale}(2009)}]{Doff:2009nk}%
  \BibitemOpen
  \bibfield  {author} {\bibinfo {author} {\bibfnamefont {A.}~\bibnamefont
  {Doff}}\ and\ \bibinfo {author} {\bibfnamefont {A.}~\bibnamefont {Natale}},\
  }\href {\doibase 10.1016/j.physletb.2009.05.045} {\bibfield  {journal}
  {\bibinfo  {journal} {Phys.Lett.}\ }\textbf {\bibinfo {volume} {B677}},\
  \bibinfo {pages} {301} (\bibinfo {year} {2009})},\ \Eprint
  {http://arxiv.org/abs/0902.2379} {arXiv:0902.2379 [hep-ph]} \BibitemShut
  {NoStop}%
%%CITATION = ARXIV:0902.2379;%%
\bibitem [{\citenamefont {Doff}\ \emph {et~al.}(2009)\citenamefont {Doff},
  \citenamefont {Natale},\ and\ \citenamefont {Rodrigues~da
  Silva}}]{Doff:2009kq}%
  \BibitemOpen
  \bibfield  {author} {\bibinfo {author} {\bibfnamefont {A.}~\bibnamefont
  {Doff}}, \bibinfo {author} {\bibfnamefont {A.}~\bibnamefont {Natale}}, \ and\
  \bibinfo {author} {\bibfnamefont {P.}~\bibnamefont {Rodrigues~da Silva}},\
  }\href {\doibase 10.1103/PhysRevD.80.055005} {\bibfield  {journal} {\bibinfo
  {journal} {Phys.Rev.}\ }\textbf {\bibinfo {volume} {D80}},\ \bibinfo {pages}
  {055005} (\bibinfo {year} {2009})},\ \Eprint {http://arxiv.org/abs/0905.2981}
  {arXiv:0905.2981 [hep-ph]} \BibitemShut {NoStop}%
%%CITATION = ARXIV:0905.2981;%%
\bibitem [{\citenamefont {Doff}\ and\ \citenamefont
  {Natale}(2010)}]{Doff:2009na}%
  \BibitemOpen
  \bibfield  {author} {\bibinfo {author} {\bibfnamefont {A.}~\bibnamefont
  {Doff}}\ and\ \bibinfo {author} {\bibfnamefont {A.}~\bibnamefont {Natale}},\
  }\href {\doibase 10.1103/PhysRevD.81.095014} {\bibfield  {journal} {\bibinfo
  {journal} {Phys.Rev.}\ }\textbf {\bibinfo {volume} {D81}},\ \bibinfo {pages}
  {095014} (\bibinfo {year} {2010})},\ \Eprint {http://arxiv.org/abs/0912.1003}
  {arXiv:0912.1003 [hep-ph]} \BibitemShut {NoStop}%
%%CITATION = ARXIV:0912.1003;%%
\bibitem [{\citenamefont {Sannino}\ and\ \citenamefont
  {Shifman}(2004)}]{Sannino:2003xe}%
  \BibitemOpen
  \bibfield  {author} {\bibinfo {author} {\bibfnamefont {F.}~\bibnamefont
  {Sannino}}\ and\ \bibinfo {author} {\bibfnamefont {M.}~\bibnamefont
  {Shifman}},\ }\href {\doibase 10.1103/PhysRevD.69.125004} {\bibfield
  {journal} {\bibinfo  {journal} {Phys.Rev.}\ }\textbf {\bibinfo {volume}
  {D69}},\ \bibinfo {pages} {125004} (\bibinfo {year} {2004})},\ \Eprint
  {http://arxiv.org/abs/hep-th/0309252} {arXiv:hep-th/0309252 [hep-th]}
  \BibitemShut {NoStop}%
%%CITATION = HEP-TH/0309252;%%
\bibitem [{\citenamefont {Hong}\ \emph {et~al.}(2004)\citenamefont {Hong},
  \citenamefont {Hsu},\ and\ \citenamefont {Sannino}}]{Hong:2004td}%
  \BibitemOpen
  \bibfield  {author} {\bibinfo {author} {\bibfnamefont {D.~K.}\ \bibnamefont
  {Hong}}, \bibinfo {author} {\bibfnamefont {S.~D.}\ \bibnamefont {Hsu}}, \
  and\ \bibinfo {author} {\bibfnamefont {F.}~\bibnamefont {Sannino}},\ }\href
  {\doibase 10.1016/j.physletb.2004.07.007} {\bibfield  {journal} {\bibinfo
  {journal} {Phys.Lett.}\ }\textbf {\bibinfo {volume} {B597}},\ \bibinfo
  {pages} {89} (\bibinfo {year} {2004})},\ \Eprint
  {http://arxiv.org/abs/hep-ph/0406200} {arXiv:hep-ph/0406200 [hep-ph]}
  \BibitemShut {NoStop}%
%%CITATION = HEP-PH/0406200;%%
\bibitem [{\citenamefont {Grinstein}\ and\ \citenamefont
  {Uttayarat}(2011)}]{Grinstein:2011dq}%
  \BibitemOpen
  \bibfield  {author} {\bibinfo {author} {\bibfnamefont {B.}~\bibnamefont
  {Grinstein}}\ and\ \bibinfo {author} {\bibfnamefont {P.}~\bibnamefont
  {Uttayarat}},\ }\href {\doibase 10.1007/JHEP07(2011)038} {\bibfield
  {journal} {\bibinfo  {journal} {JHEP}\ }\textbf {\bibinfo {volume} {1107}},\
  \bibinfo {pages} {038} (\bibinfo {year} {2011})},\ \Eprint
  {http://arxiv.org/abs/1105.2370} {arXiv:1105.2370 [hep-ph]} \BibitemShut
  {NoStop}%
%%CITATION = ARXIV:1105.2370;%%
\bibitem [{\citenamefont {Antipin}\ \emph
  {et~al.}(2012{\natexlab{c}})\citenamefont {Antipin}, \citenamefont {Mojaza},\
  and\ \citenamefont {Sannino}}]{Antipin:2011aa}%
  \BibitemOpen
  \bibfield  {author} {\bibinfo {author} {\bibfnamefont {O.}~\bibnamefont
  {Antipin}}, \bibinfo {author} {\bibfnamefont {M.}~\bibnamefont {Mojaza}}, \
  and\ \bibinfo {author} {\bibfnamefont {F.}~\bibnamefont {Sannino}},\ }\href
  {\doibase 10.1016/j.physletb.2012.04.050} {\bibfield  {journal} {\bibinfo
  {journal} {Phys.Lett.}\ }\textbf {\bibinfo {volume} {B712}},\ \bibinfo
  {pages} {119} (\bibinfo {year} {2012}{\natexlab{c}})},\ \Eprint
  {http://arxiv.org/abs/1107.2932} {arXiv:1107.2932 [hep-ph]} \BibitemShut
  {NoStop}%
%%CITATION = ARXIV:1107.2932;%%
\bibitem [{\citenamefont {Kennedy}\ and\ \citenamefont
  {Lynn}(1989)}]{Kennedy:1988sn}%
  \BibitemOpen
  \bibfield  {author} {\bibinfo {author} {\bibfnamefont {D.}~\bibnamefont
  {Kennedy}}\ and\ \bibinfo {author} {\bibfnamefont {B.}~\bibnamefont {Lynn}},\
  }\href {\doibase 10.1016/0550-3213(89)90483-5} {\bibfield  {journal}
  {\bibinfo  {journal} {Nucl.Phys.}\ }\textbf {\bibinfo {volume} {B322}},\
  \bibinfo {pages} {1} (\bibinfo {year} {1989})}\BibitemShut {NoStop}%
%%CITATION = NUPHA,B322,1;%%
\bibitem [{\citenamefont {Peskin}\ and\ \citenamefont
  {Takeuchi}(1990)}]{Peskin:1990zt}%
  \BibitemOpen
  \bibfield  {author} {\bibinfo {author} {\bibfnamefont {M.~E.}\ \bibnamefont
  {Peskin}}\ and\ \bibinfo {author} {\bibfnamefont {T.}~\bibnamefont
  {Takeuchi}},\ }\href {\doibase 10.1103/PhysRevLett.65.964} {\bibfield
  {journal} {\bibinfo  {journal} {Phys.Rev.Lett.}\ }\textbf {\bibinfo {volume}
  {65}},\ \bibinfo {pages} {964} (\bibinfo {year} {1990})}\BibitemShut
  {NoStop}%
%%CITATION = PRLTA,65,964;%%
\bibitem [{\citenamefont {Baak}\ \emph {et~al.}(2012)\citenamefont {Baak},
  \citenamefont {Goebel}, \citenamefont {Haller}, \citenamefont {Hoecker},
  \citenamefont {Kennedy} \emph {et~al.}}]{Baak:2012kk}%
  \BibitemOpen
  \bibfield  {author} {\bibinfo {author} {\bibfnamefont {M.}~\bibnamefont
  {Baak}}, \bibinfo {author} {\bibfnamefont {M.}~\bibnamefont {Goebel}},
  \bibinfo {author} {\bibfnamefont {J.}~\bibnamefont {Haller}}, \bibinfo
  {author} {\bibfnamefont {A.}~\bibnamefont {Hoecker}}, \bibinfo {author}
  {\bibfnamefont {D.}~\bibnamefont {Kennedy}},  \emph {et~al.},\ }\href
  {\doibase 10.1140/epjc/s10052-012-2205-9} {\bibfield  {journal} {\bibinfo
  {journal} {Eur.Phys.J.}\ }\textbf {\bibinfo {volume} {C72}},\ \bibinfo
  {pages} {2205} (\bibinfo {year} {2012})},\ \Eprint
  {http://arxiv.org/abs/1209.2716} {arXiv:1209.2716 [hep-ph]} \BibitemShut
  {NoStop}%
%%CITATION = ARXIV:1209.2716;%%
\bibitem [{\citenamefont {Sannino}\ and\ \citenamefont
  {Tuominen}(2005)}]{Sannino:2004qp}%
  \BibitemOpen
  \bibfield  {author} {\bibinfo {author} {\bibfnamefont {F.}~\bibnamefont
  {Sannino}}\ and\ \bibinfo {author} {\bibfnamefont {K.}~\bibnamefont
  {Tuominen}},\ }\href {\doibase 10.1103/PhysRevD.71.051901} {\bibfield
  {journal} {\bibinfo  {journal} {Phys.Rev.}\ }\textbf {\bibinfo {volume}
  {D71}},\ \bibinfo {pages} {051901} (\bibinfo {year} {2005})},\ \Eprint
  {http://arxiv.org/abs/hep-ph/0405209} {arXiv:hep-ph/0405209 [hep-ph]}
  \BibitemShut {NoStop}%
%%CITATION = HEP-PH/0405209;%%
\bibitem [{\citenamefont {Dietrich}\ \emph {et~al.}(2006)\citenamefont
  {Dietrich}, \citenamefont {Sannino},\ and\ \citenamefont
  {Tuominen}}]{Dietrich:2005wk}%
  \BibitemOpen
  \bibfield  {author} {\bibinfo {author} {\bibfnamefont {D.~D.}\ \bibnamefont
  {Dietrich}}, \bibinfo {author} {\bibfnamefont {F.}~\bibnamefont {Sannino}}, \
  and\ \bibinfo {author} {\bibfnamefont {K.}~\bibnamefont {Tuominen}},\ }\href
  {\doibase 10.1103/PhysRevD.73.037701} {\bibfield  {journal} {\bibinfo
  {journal} {Phys.Rev.}\ }\textbf {\bibinfo {volume} {D73}},\ \bibinfo {pages}
  {037701} (\bibinfo {year} {2006})},\ \Eprint
  {http://arxiv.org/abs/hep-ph/0510217} {arXiv:hep-ph/0510217 [hep-ph]}
  \BibitemShut {NoStop}%
%%CITATION = HEP-PH/0510217;%%
\bibitem [{\citenamefont {Foadi}\ \emph {et~al.}(2007)\citenamefont {Foadi},
  \citenamefont {Frandsen}, \citenamefont {Ryttov},\ and\ \citenamefont
  {Sannino}}]{Foadi:2007ue}%
  \BibitemOpen
  \bibfield  {author} {\bibinfo {author} {\bibfnamefont {R.}~\bibnamefont
  {Foadi}}, \bibinfo {author} {\bibfnamefont {M.~T.}\ \bibnamefont {Frandsen}},
  \bibinfo {author} {\bibfnamefont {T.~A.}\ \bibnamefont {Ryttov}}, \ and\
  \bibinfo {author} {\bibfnamefont {F.}~\bibnamefont {Sannino}},\ }\href
  {\doibase 10.1103/PhysRevD.76.055005} {\bibfield  {journal} {\bibinfo
  {journal} {Phys.Rev.}\ }\textbf {\bibinfo {volume} {D76}},\ \bibinfo {pages}
  {055005} (\bibinfo {year} {2007})},\ \Eprint {http://arxiv.org/abs/0706.1696}
  {arXiv:0706.1696 [hep-ph]} \BibitemShut {NoStop}%
%%CITATION = ARXIV:0706.1696;%%
\bibitem [{\citenamefont {Foadi}\ \emph {et~al.}(2008)\citenamefont {Foadi},
  \citenamefont {Frandsen},\ and\ \citenamefont {Sannino}}]{Foadi:2007se}%
  \BibitemOpen
  \bibfield  {author} {\bibinfo {author} {\bibfnamefont {R.}~\bibnamefont
  {Foadi}}, \bibinfo {author} {\bibfnamefont {M.~T.}\ \bibnamefont {Frandsen}},
  \ and\ \bibinfo {author} {\bibfnamefont {F.}~\bibnamefont {Sannino}},\ }\href
  {\doibase 10.1103/PhysRevD.77.097702} {\bibfield  {journal} {\bibinfo
  {journal} {Phys.Rev.}\ }\textbf {\bibinfo {volume} {D77}},\ \bibinfo {pages}
  {097702} (\bibinfo {year} {2008})},\ \Eprint {http://arxiv.org/abs/0712.1948}
  {arXiv:0712.1948 [hep-ph]} \BibitemShut {NoStop}%
%%CITATION = ARXIV:0712.1948;%%
\bibitem [{\citenamefont {Ryttov}\ and\ \citenamefont
  {Sannino}(2008)}]{Ryttov:2008xe}%
  \BibitemOpen
  \bibfield  {author} {\bibinfo {author} {\bibfnamefont {T.~A.}\ \bibnamefont
  {Ryttov}}\ and\ \bibinfo {author} {\bibfnamefont {F.}~\bibnamefont
  {Sannino}},\ }\href {\doibase 10.1103/PhysRevD.78.115010} {\bibfield
  {journal} {\bibinfo  {journal} {Phys.Rev.}\ }\textbf {\bibinfo {volume}
  {D78}},\ \bibinfo {pages} {115010} (\bibinfo {year} {2008})},\ \Eprint
  {http://arxiv.org/abs/0809.0713} {arXiv:0809.0713 [hep-ph]} \BibitemShut
  {NoStop}%
%%CITATION = ARXIV:0809.0713;%%
\bibitem [{\citenamefont {Pich}\ \emph {et~al.}(2012)\citenamefont {Pich},
  \citenamefont {Rosell},\ and\ \citenamefont {Sanz-Cillero}}]{Pich:2012dv}%
  \BibitemOpen
  \bibfield  {author} {\bibinfo {author} {\bibfnamefont {A.}~\bibnamefont
  {Pich}}, \bibinfo {author} {\bibfnamefont {I.}~\bibnamefont {Rosell}}, \ and\
  \bibinfo {author} {\bibfnamefont {J.~J.}\ \bibnamefont {Sanz-Cillero}},\
  }\href@noop {} {\  (\bibinfo {year} {2012})},\ \Eprint
  {http://arxiv.org/abs/1212.6769} {arXiv:1212.6769 [hep-ph]} \BibitemShut
  {NoStop}%
%%CITATION = ARXIV:1212.6769;%%
\bibitem [{\citenamefont {Appelquist}\ \emph {et~al.}(1986)\citenamefont
  {Appelquist}, \citenamefont {Karabali},\ and\ \citenamefont
  {Wijewardhana}}]{Appelquist:1986an}%
  \BibitemOpen
  \bibfield  {author} {\bibinfo {author} {\bibfnamefont {T.~W.}\ \bibnamefont
  {Appelquist}}, \bibinfo {author} {\bibfnamefont {D.}~\bibnamefont
  {Karabali}}, \ and\ \bibinfo {author} {\bibfnamefont {L.}~\bibnamefont
  {Wijewardhana}},\ }\href {\doibase 10.1103/PhysRevLett.57.957} {\bibfield
  {journal} {\bibinfo  {journal} {Phys.Rev.Lett.}\ }\textbf {\bibinfo {volume}
  {57}},\ \bibinfo {pages} {957} (\bibinfo {year} {1986})}\BibitemShut
  {NoStop}%
%%CITATION = PRLTA,57,957;%%
\bibitem [{\citenamefont {Foadi}\ and\ \citenamefont
  {Sannino}(2008)}]{Foadi:2008ci}%
  \BibitemOpen
  \bibfield  {author} {\bibinfo {author} {\bibfnamefont {R.}~\bibnamefont
  {Foadi}}\ and\ \bibinfo {author} {\bibfnamefont {F.}~\bibnamefont
  {Sannino}},\ }\href {\doibase 10.1103/PhysRevD.78.037701} {\bibfield
  {journal} {\bibinfo  {journal} {Phys.Rev.}\ }\textbf {\bibinfo {volume}
  {D78}},\ \bibinfo {pages} {037701} (\bibinfo {year} {2008})},\ \Eprint
  {http://arxiv.org/abs/0801.0663} {arXiv:0801.0663 [hep-ph]} \BibitemShut
  {NoStop}%
%%CITATION = ARXIV:0801.0663;%%
\bibitem [{\citenamefont {Fodor}\ \emph
  {et~al.}(2012{\natexlab{a}})\citenamefont {Fodor}, \citenamefont {Holland},
  \citenamefont {Kuti}, \citenamefont {Nogradi}, \citenamefont {Schroeder}
  \emph {et~al.}}]{Fodor:2012ty}%
  \BibitemOpen
  \bibfield  {author} {\bibinfo {author} {\bibfnamefont {Z.}~\bibnamefont
  {Fodor}}, \bibinfo {author} {\bibfnamefont {K.}~\bibnamefont {Holland}},
  \bibinfo {author} {\bibfnamefont {J.}~\bibnamefont {Kuti}}, \bibinfo {author}
  {\bibfnamefont {D.}~\bibnamefont {Nogradi}}, \bibinfo {author} {\bibfnamefont
  {C.}~\bibnamefont {Schroeder}},  \emph {et~al.},\ }\href {\doibase
  10.1016/j.physletb.2012.10.079} {\bibfield  {journal} {\bibinfo  {journal}
  {Phys.Lett.}\ }\textbf {\bibinfo {volume} {B718}},\ \bibinfo {pages} {657}
  (\bibinfo {year} {2012}{\natexlab{a}})},\ \Eprint
  {http://arxiv.org/abs/1209.0391} {arXiv:1209.0391 [hep-lat]} \BibitemShut
  {NoStop}%
%%CITATION = ARXIV:1209.0391;%%
\bibitem [{\citenamefont {Fodor}\ \emph
  {et~al.}(2012{\natexlab{b}})\citenamefont {Fodor}, \citenamefont {Holland},
  \citenamefont {Kuti}, \citenamefont {Nogradi}, \citenamefont {Schroeder}
  \emph {et~al.}}]{Fodor:2012ni}%
  \BibitemOpen
  \bibfield  {author} {\bibinfo {author} {\bibfnamefont {Z.}~\bibnamefont
  {Fodor}}, \bibinfo {author} {\bibfnamefont {K.}~\bibnamefont {Holland}},
  \bibinfo {author} {\bibfnamefont {J.}~\bibnamefont {Kuti}}, \bibinfo {author}
  {\bibfnamefont {D.}~\bibnamefont {Nogradi}}, \bibinfo {author} {\bibfnamefont
  {C.}~\bibnamefont {Schroeder}},  \emph {et~al.},\ }\href@noop {} {\bibfield
  {journal} {\bibinfo  {journal} {PoS}\ }\textbf {\bibinfo {volume}
  {LATTICE2012}},\ \bibinfo {pages} {024} (\bibinfo {year}
  {2012}{\natexlab{b}})},\ \Eprint {http://arxiv.org/abs/1211.6164}
  {arXiv:1211.6164 [hep-lat]} \BibitemShut {NoStop}%
%%CITATION = ARXIV:1211.6164;%%
\end{thebibliography}%

\end{document}